\documentclass[prx,showpacs,superscriptaddress,amssymb,twocolumn,notitlepage,nofootinbib,longbibliography,preprintnumbers]{revtex4-2}
\usepackage{graphicx}
\usepackage{amsmath}

\usepackage{amssymb}
\usepackage{hyperref}
\hypersetup{
colorlinks=true,       
linkcolor=cyan,          
citecolor=magenta,        
filecolor=magenta,      
urlcolor=cyan,           
runcolor=cyan
}

\usepackage[dvipsnames]{xcolor}
\usepackage{datetime2}
\usepackage{soul}

\usepackage{bm}
\usepackage{threeparttable}
\usepackage{subfigure}
\usepackage{upgreek }
\usepackage{marginnote}
\usepackage{cancel}

\newcommand{\beq}{\begin{equation}}
\newcommand{\eeq}{\end{equation}}
\newcommand{\beqnn}{\begin{equation*}}
\newcommand{\eeqnn}{\end{equation*}}
\newcommand{\bea}{\begin{eqnarray}}
\newcommand{\eea}{\end{eqnarray}}
\newcommand{\beann}{\begin{eqnarray*}}
\newcommand{\eeann}{\end{eqnarray*}}
\newcommand{\bes} {\begin{subequations}}
\newcommand{\ees} {\end{subequations}}

\newcommand{\ignore}[1]{}

\DeclareRobustCommand{\trace}[1]{\ensuremath{\mbox{Tr}\left[#1\right]}}

\begin{document}

\title{Hamiltonian Learning using Machine Learning Models Trained with Continuous Measurements}

\author{Kris Tucker}
\email{kristopher.tucker@colorado.edu}
\affiliation{Department of Applied Math, University of Colorado Boulder, USA}

\author{Amit Kiran Rege}
\email{Amit.Rege@colorado.edu}
\affiliation{Department of Computer Science, University of Colorado Boulder, USA}

\author{Conor Smith}
\affiliation{Center for Quantum Information and Control, University of New Mexico, University of New Mexico, Albuquerque, NM, USA}
\affiliation{Department of Electrical and Computer Engineering, University of New Mexico, Albuquerque, NM, USA}

\author{Claire Monteleoni}
\affiliation{Department of Computer Science, University of Colorado Boulder, USA}
\affiliation{INRIA Centre de Recherche de Paris, France}

\author{Tameem Albash}
\affiliation{Center for Computing Research, Sandia National Laboratories, Albuquerque, New Mexico 87185, USA}

\begin{abstract}
    We build upon recent work on using Machine Learning models to estimate Hamiltonian parameters using continuous weak measurement of qubits as input. We consider two settings for the training of our model: (1) supervised learning where the weak measurement training record can be labeled with known Hamiltonian parameters, and (2) unsupervised learning where no labels are available. The first has the advantage of not requiring an explicit representation of the quantum state, thus potentially scaling very favorably to a larger number of qubits.  The second requires the implementation of a physical model to map the Hamiltonian parameters to a measurement record, which we implement using an integrator of the physical model with a recurrent neural network to provide a model-free correction at every time step to account for small effects not captured by the physical model. We test our construction on a system of two qubits and demonstrate accurate prediction of multiple physical parameters in both the supervised and unsupervised contexts. We demonstrate that the model benefits from larger training sets establishing that it is in fact ``learning,'' and we show robustness to errors in the assumed physical model by achieving accurate parameter estimation in the presence of unanticipated single particle relaxation.
\end{abstract}
\maketitle

\section{Introduction}

As the scale, complexity, and availability of quantum devices continues to grow over the coming decades \cite{preskill_nisq}, the ability to accurately characterize device parameters, especially unwanted effects, will become ever-more important. For example, non-local errors such as crosstalk in large systems  \cite{crosstalk_sarovar, crosstalk_lidar} can be problematic for quantum error correction methods \cite{Kitaev1997,Aharonov1997,Preskill1998,Knill1998,Gottesman1998}. Detection of these unwanted effects is a critical first step in mitigating them, with efforts to do so complicated by the exponential scaling of the dimension of the Hilbert space with system size and the challenges of modeling increasingly complex systems.

Recent advances in the application of machine learning (ML) tools to quantum systems have shown promise in overcoming some of these difficulties. Scalable ML models have been used to represent quantum states for state tomography \cite{carleo_troyer,carleo_2018,qst_1810_10584,qst_vae,qst_tutorial,qpt} and evolution \cite{qpt,nm_rnn}, for learning unknown dynamics \cite{sindy,sindy_champion,piml_encode}, as well as for learning device characterization from measurements \cite{PRXQuantum.2.040355,siddiqi_rnn,siddiqi_rnn2,zan_trans,castelano_piml}. Increasingly, a priori knowledge of a physical system has been combined with ML models to improve accuracy and interpretability \cite{Raissi_PIML2}. The model-free nature of many ML solutions has made them scalable and robust to many of the common pitfalls of physical models, such as non-Markovian dynamics \cite{PRXQuantum.2.040355,siddiqi_rnn,siddiqi_rnn2,nm_rnn}. There are trade-offs, however, as more abstract models are generally less interpretable, which limits the physical insights that can be gleaned from them. Finding a balance between these two competing properties, representability and interpretability, thus becomes an important challenge.

In this work, we build on recent advances in device characterization applying ML models to the continuous measurement of qubits \cite{siddiqi_rnn, siddiqi_rnn2, PRXQuantum.2.040355}. 
In contrast with the parameter estimation approach of Ref.~\cite{PRXQuantum.2.040355}, where a stochastic master equation is used as a trainable model, our ML model learns a direct map from continuous measurement inputs to the system parameters of interest. In the case where the measurement records are associated with known device parameters (the case of supervised learning), this approach has two advantages. First, the model is completely independent of any state representation (in contrast with approaches like Ref.~\cite{zan_trans}). The measurements need only provide enough information to estimate the parameters of interest and not the complete quantum state, potentially freeing it from the curse of dimensionality that comes from the exponential scaling of the dimension of the Hilbert space and thus making it more scalable. 

Second, once the ML model is trained, parameter estimation can be performed quickly even for systems with parameters not seen during training. It can also benefit from having been provided with many examples of noisy measurement records associated with known true values in the training set. This makes it possible for the model to learn to distinguish between those features of measurement records that vary with parameters and those that are uncorrelated noise, thus potentially requiring fewer measurements for estimation.

In the event that device parameters are not provided along with measurement records, an unsupervised approach comparing the input to a measurement record reconstructed from parameter estimates can be used. This requires a map from parameter estimates to measurement output, which is accomplished by adding a layer combining an integrator of the physical model with a recurrent neural network (RNN) to provide a model-free correction at every time step to account for unanticipated effects, as long as they are consistent and small relative to the dynamics driven by the parameters of interest. This employs the capabilities of neural ordinary differential equations found in other works \cite{neural_ode,odernn,nm_rnn} but enhanced to provide a completely model-free correction not bound by assumptions of linearity or Markovianity,  with a projection step to return the corrected state to the manifold of physical density operators \cite{proj_density}. The output of the model is then the estimated solution to the unconditioned master equation with learned corrections for effects beyond the master equation, which can be compared to the measurement records to update model parameters. While this case is not completely model-free, it also does not make rigid assumptions about the dynamics and remains capable of accurate parameter estimation in the presence of unanticipated effects that would otherwise severely impact accuracy. The ML model is not burdened with learning all of the physics; it just has to correct for small effects the model may have missed, an approach known as discrepancy modeling \cite{disc,disc2}. 

The fully unsupervised model can be viewed as a denoising autoencoder (DNA) \cite{goodfellow} where the physical parameters being estimated take on the role of the latent space, with interpretability enforced by the presence of the physical model in the decoder. The encoder is the map from measurement records to parameters, and is the desired product of the training to be used for fast and accurate parameter estimation.

While the unsupervised model clearly has applications for parameter estimation in cases where nothing is known about the parameters of interest, the supervised approach using just the encoder could still find applications for systems where parameters are known at the time training data is generated, but a prediction routine is still required in other circumstances, as would be the case for detecting drift away from device calibration over time. We apply both approaches to the specific problem of learning one- and two-body Hamiltonian parameters in a two qubit system. 

Finally, we contrast our work with recent methods for Hamiltonian learning based on Gibbs state measurements or real-time evolution \cite{zan_trans, Haah2024, Anshu2021, Anshu2024, PhysRevLett.130.200403, arxiv2307.04690, arxiv2304.07172, PhysRevLett.122.020504, Bairey_2020}. In our work the model is trained and performs predictions based on continuous measurements modeled by a stochastic master equation rather than Hamiltonian or deterministic master equation evolution combined with strong measurements. Our approach that is applicable to both Hamiltonian and Lindblad parameters requires that the parameters to be learned have an observable impact on the continuous measurements, and it does not use the preparation of a steady state or energy eigenstate~\cite{zan_trans,Anshu2021,Bairey_2020} or changing the dynamics by adaptively changing the Hamiltonian~\cite{arxiv2304.07172} or introducing additional dynamics~\cite{PhysRevLett.130.200403,arxiv2307.04690}. It is likely adaptive approaches could enhance our learning as well. Finally, the discrepancy modeling in our approach is unique in that it allows for the combination of knowledge about the physical system with model-free machine learning to enhance estimation accuracy and ML model interpretability.

The manuscript is organized as follows. In Sec.~\ref{sec:PhysicalSystem}, we describe the physical system of interest to us, which will be given by two qubits that are weakly measured. In Sec.~\ref{sec:ML}, we describe our ML models and their supervised and unsupervised training. In Sec.~\ref{sec_results}, we show results for the performance of the models.  In Sec.~\ref{sec:conclusions}, we summarize our results and discuss possible future extensions of this work.

\section{Physical System} \label{sec:PhysicalSystem}

\begin{figure}[hb!]
    \centering
    \includegraphics[width=7.5cm]{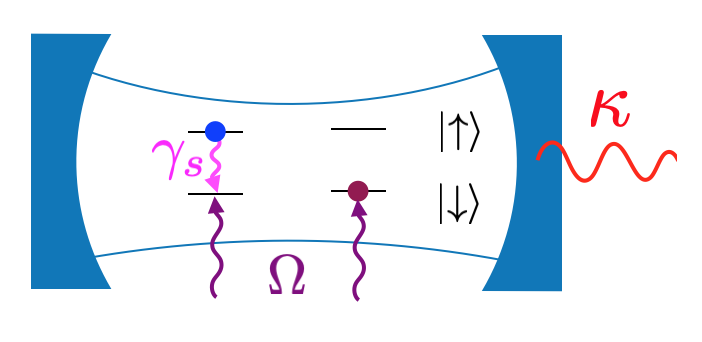}
    \caption{The physical system of two qubits in a microwave cavity subject to a Rabi drive $\Omega$, single particle relaxation with rate $\gamma_s$, and dephasing due to weak measurement back-action with rate $\kappa$.}
    \label{fig:phys_system}
\end{figure}

Our physical system of interest is that of two qubits with fixed position in a microwave cavity as illustrated in Fig.~\ref{fig:phys_system}. A single common mode of the cavity is coupled to the computational degree of freedom of the qubits, and they are coherently driven on resonance with a Rabi drive of frequency $\Omega$. A two-qubit interaction term is present with magnitude $\epsilon$. A weak measurement tone \cite{cont_meas,brun} is applied to the cavity to probe the qubit state in one of the $\{X,Y,Z\}$ directions for each qubit, with a measurement back-action dephasing rate of $\kappa$. Upon adiabatic elimination of the cavity mode \cite{Haake_1971}, which we assume has dynamics evolving at a rate much faster than time scales relevant to the qubits, we can describe the system with the stochastic master equation (SME) \cite{circuit_qed,circuit_qed_traj}
\bes \label{eqt:me}
\begin{align} 
    d\rho &= -i[H,\rho] dt \nonumber \\
    & + \sum_{i=1}^2 \mathcal{D}[L_i](\rho) dt + \sum_{i=1}^2 \sqrt{\frac{\eta}{2}}\mathcal{H}[L_i](\rho) dW_t^{(i)},
    \label{eqn:sme} \\
    H &= \sum_{i=1}^2\frac{\Omega}{2} X_i + \epsilon Z_1 Z_2,
    \label{eqn:ham}
\end{align}
\ees
where $\{X_i,Y_i,Z_i\}$ is the set of Pauli operators for qubit $i \in \left\{ 1, 2 \right\}$, $\mathcal{D}[L](\rho) = L\rho L^\dagger - \frac{1}{2}\left\{L^\dagger L\rho + \rho L^\dagger L\right\}$ is the Lindblad super-operator, $\mathcal{H}[L](\rho) = L\rho + \rho L^\dagger - \rho \trace{\rho\left(L + L^\dagger\right)}$ is the measurement super-operator, $\eta$ is the efficiency of the measurement, and $L_i = \sqrt{\kappa}C_i,$ where $C_i \in \{X_i,Y_i,Z_i\}$ is the weak measurement operator. Here we have suppressed the dependence on $t$ for $\rho$, and we have adopted the normalization convention $\hbar = 1$.

The stochastic differential equations for the measurement records are given by
\begin{equation}
    dr_i = \sqrt{\frac{\eta}{2}}\trace{\rho\left(L_i + L_i^\dagger\right)} dt + dW_t^{(i)} 
    \label{eqn:volt}
\end{equation}
where $r_i$ is the weak measurement record for qubit $i\in \left\{ 1, 2 \right\}$ and the independent Wiener increments $dW_t^{(i)}$ are the same as those appearing in Eq.~\eqref{eqn:sme}. This is comparable to the system studied in Ref.~\cite{PRXQuantum.2.040355}, except generalized to two qubits and with the addition of the two-qubit interaction term in Eq.~\eqref{eqn:ham}. We also consider cases where the weak measurement operator $L_i$ is different for each qubit.

The unconditioned master equation is given by averaging over all possible trajectories of the Wiener processes and is given by:
\bes \label{eqt:uncme}
\begin{align} 
    d\overline{\rho} &= -i[H,\overline{\rho}] dt + \sum_{i=1}^2 \mathcal{D}[L_i](\overline{\rho}) dt \label{eqt:uncme_a} \\
    d \overline{r_i} &= \sqrt{\frac{\eta}{2}}\trace{\overline{\rho}\left(L_i + L_i^\dagger\right)} dt 
    \label{eqt:uncme_b}
\end{align}
\ees

\section{Machine Learning Model} \label{sec:ML}

\subsection{Preliminaries}
We briefly review the kinds of architectures used in this work. For a more detailed overview, we refer the reader to Appendix \ref{sec:rnn}.

Recurrent Neural Networks (RNNs) \cite{rumelhart:errorpropnonote} are a class of artificial neural networks designed to recognize patterns in sequences of data. Unlike traditional neural networks, which assume all inputs (and outputs) are independent of each other, RNNs are characterized by their ability to retain information from previous inputs in the sequence through internal memory, making them ideal for tasks involving sequential data.

Traditional RNNs struggle to learn and retain information over long sequences due to problems like the contribution of information decaying over time \cite{Hochreiter:01book}. This makes it hard for the network to maintain long-term dependencies. Long Short-Term Memory (LSTM) networks \cite{lstm} are a type of RNN architecture designed to address the issue of long-term dependencies in sequence data. In this work, we utilize LSTM-based models in order to model trajectories of weak measurements.

We now describe a common design recipe used in tasks performing unsupervised translation of data from one domain to another, often referred to, collectively, as the family of encoder-decoder models. These are also called sequence-to-sequence models when working with sequential data. Such models produce an arbitrary length, context-dependent sequence, given an input sequence. The length of the input sequence need not match the output sequence, and individual elements of the input sequence need not have one-to-one correspondence with individual elements of the output sequence. 

\begin{figure}[th!]
    \centering
    \includegraphics[width=0.5\textwidth]{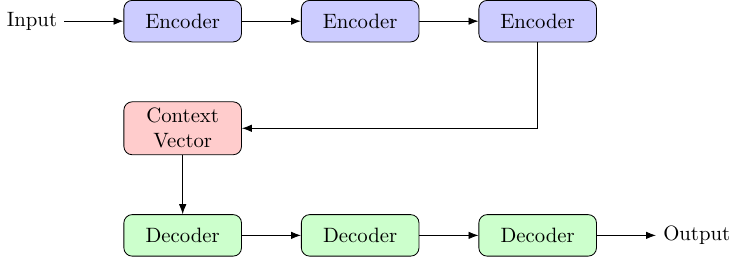}
    \caption{A high-level description of the encoder-decoder design recipe.}
    \label{fig:seq2seq}
\end{figure}

The key idea is to use the encoder network to obtain a compact representation of the input sequence (often called the context vector). This representation is then passed to the decoder network to produce an output sequence by augmenting it to the hidden state in the decoder. This is an extremely general framework and potentially any neural network can be used in such an arrangement (see Fig \ref{fig:seq2seq}).

In our unsupervised setup, the encoder is an LSTM that returns the master equation parameters as the context. The decoder is a numerical ODE integrator implemented as an RNN.

\subsection{Model Description}

\begin{figure}[t!]
    \centering
    \includegraphics[width=7.5cm]{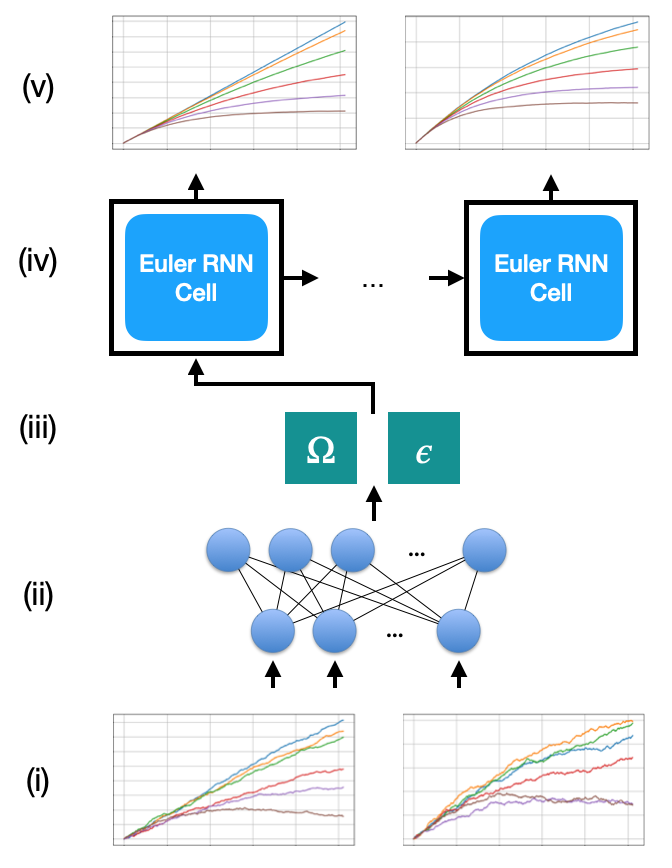}
    \caption{Diagram of the full ML model. Voltage records are measured from the cavity system producing noisy averaged voltage records (i), these are sent through a neural network encoder (ii) producing system parameters as output (iii). These parameters are used by the flex integrator decoder (iv) to produce noise-free voltage estimates (v).}
    \label{fig:ml_model}
\end{figure}

Our objective is to train a ML model to estimate the physical master equation parameters in Eq.~\eqref{eqt:me} by observing the weak measurement records described by Eq.~\eqref{eqn:volt}. Input data is provided for training of the ML model in the form of a collection of \emph{averaged} weak measurement records, where the average is over some number of records $d$ to reduce the effects of diffusive noise. The output of the model is either (a) the physical master equation parameters of interest in the supervised case where labeled training data is available, or (b) the predicted measurement record satisfying the unconditioned master equation (Eq.~\eqref{eqt:uncme}) in the unsupervised case where true parameters are not available. The former configuration can be interpreted as an encoder mapping noisy trajectories to estimated parameter values, while the latter configuration adds a decoder that takes the estimated parameters as input and produces a predicted solution to the unconditioned master equation. The full architecture is shown in Fig.~\ref{fig:ml_model} and can be viewed as a denoising autoencoder taking noisy input trajectories and producing output with the noise removed. 

The encoder begins with an average pooling operation in the time dimension to further smooth the trend component of the voltages, followed by a long short-term memory (LSTM) layer \cite{lstm} to process the sequential input data and consolidate information from multiple qubits into a single sequence. This is followed by a feed-forward neural network of dense layers ending with the parameter layer. The LSTM and feed-forward layers comprise the neural network shown in Fig.~\ref{fig:ml_model}(ii).

\begin{figure}[ht!]
    \centering
    \includegraphics[width=7.5cm]{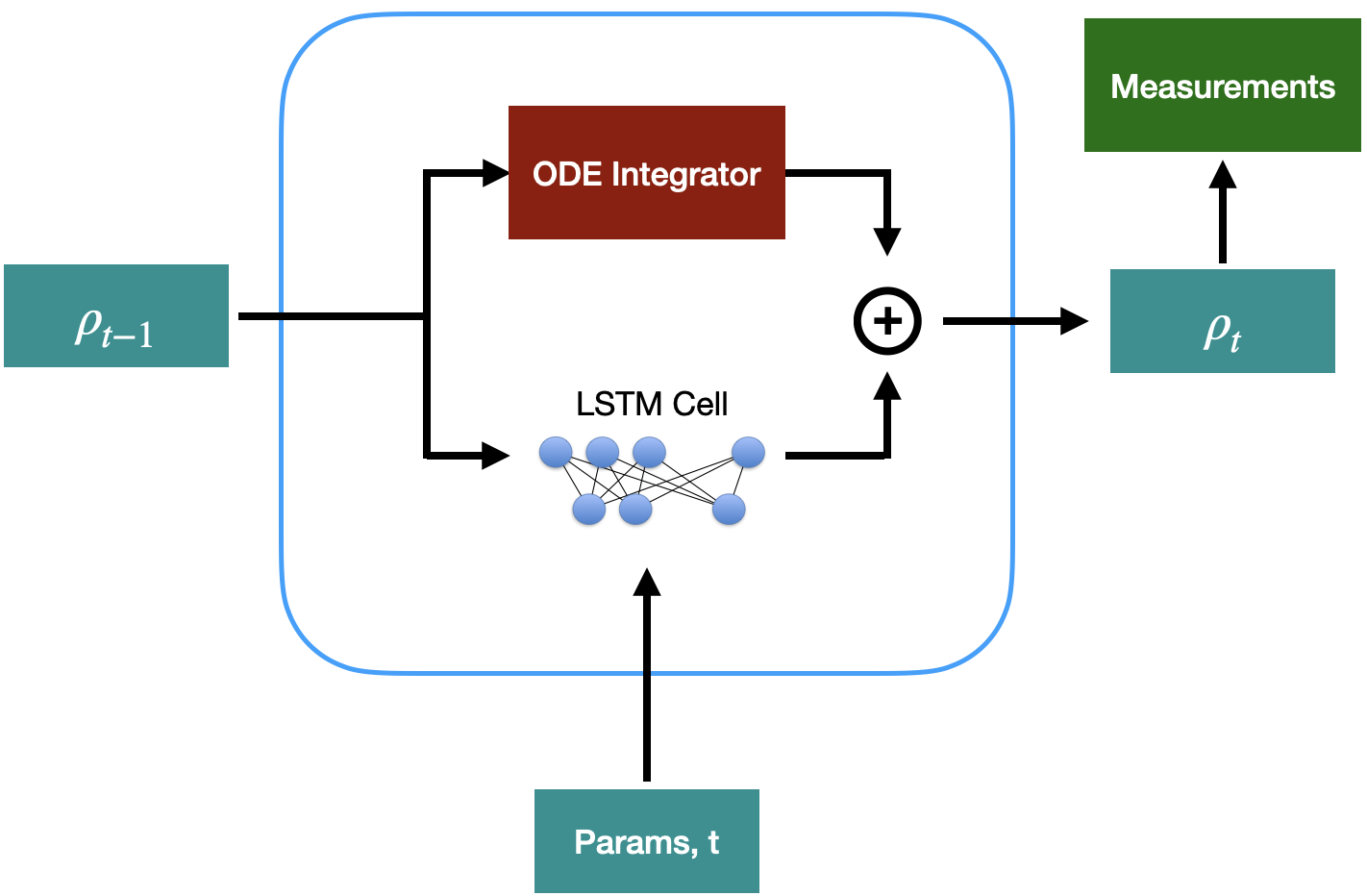}
    \caption{Details of the decoder RNN cell combining a standard ODE integrator with an LSTM update.}
    \label{fig:flex_cell}
\end{figure}

In the unsupervised case, interpretability of the latent variables as parameters is enforced by using them directly in an enhanced numerical ODE integrator in the decoder for the unconditioned master equation. The integrator is implemented as a RNN with a custom cell that combines a single step of Euler's method with a correction produced by a standard LSTM cell. This is shown in Fig.~\ref{fig:flex_cell}. The correction term is designed to compensate for unanticipated dynamics in the physical model used by the numerical integrator. The design leverages the concept of neural ODE \cite{neural_ode} or ODE-RNN \cite{odernn,odernn_latent}, as it uses a neural network to approximate corrections to the unconditioned master equation as:
\begin{equation}
    d \overline{\rho}  = f_{\text{model}}\left(t,\overline{\rho};\theta\right)dt + f_{\text{LSTM}}\left(t,\overline{\rho};\theta\right)dt,
\end{equation}
where $\theta$ is a set of ODE parameters, $f_{\text{model}}$ is the drift term of the form of Eq.~\eqref{eqt:uncme}, and $f_{\text{LSTM}}$ is an LSTM cell and as such is a non-linear, model free function of the time, state, and parameters. While we are assuming for this work, as discussed in Sec.~\ref{sec_data}, that the parameters of the measurement equation in Eq.~\eqref{eqt:uncme_b} are known with high precision, this approach could easily be extended to this equation as well.

The carry-state used in the standard LSTM architecture is passed between evaluations of $f_{\text{LSTM}}$ at each time point, which means information from the full history of states is potentially available, allowing for the modeling of non-Markovian effects 
We ensure that the output operator is a quantum state at each time step by restricting the free parameters in the correction to have a Hermitian form to enforce Hermiticity, by implementing a normalization step to ensure that the trace is preserved, and by performing an orthogonal projection of the operator back onto the state space according to the algorithm presented in Ref.~\cite{proj_density}. In this respect it is analogous to the approach used by the time-dependent variational principle (TDVP) algorithm \cite{tdvp, tdvp2} for modeling dynamics of matrix product states (MPS)~\cite{perez2006matrix, schollwock2011density}.

We emphasize that the enhanced numerical integrator is primarily relying on a known physical model given by $f_{\text{model}}$ and using the ML components in $f_{\text{LSTM}}$ only to compensate for discrepancies between this model and the experimental system, which are assumed to be small relative to the known dynamics. As such, it could be categorized along with other approaches in ML for discrepancy modeling \cite{disc,disc2} that operate under the principle that small corrections are easier to learn than the full dynamics. A similar approach is taken to modeling non-Markovian dynamics via RNN in Ref.~\cite{nm_rnn}, though our approach differs in that it combines discrepancy modeling with a completely general, non-linear correction to the state evolution based on a neural ODE. In addition, as suggested in the outlook section of Ref.~\cite{nm_rnn}, our investigation focuses on the situation where only a subset of the measurement information necessary for full state reconstruction is available, so training data volume requirements are substantially reduced.

It should be noted that while the integrator used in the decoder for this work is an Euler integrator for the unconditioned master equation, this piece is completely modular in design and can be replaced with any integrator for simulating quantum dynamics. For example, a TDVP integrator using MPS could be used for larger systems that are approximated well by a MPS ansatz, with the trainable decoder parameters accounting for errors introduced by this approach. It should also be noted that, since the output of the decoder is the solution to the unconditioned master equation, it contains an ODE rather than an SDE integrator. Once trained, however, it would be possible to switch over to an SDE integrator, for example by switching from Euler to Euler-Maruyama, by simulating noise. The model would then serve as a generative model that could be useful for simulation purposes.

\subsection{Model Interpretation}

The output of the decoder is the estimated solution to the unconditioned master equation (Eq.~\eqref{eqt:uncme_b}), and the loss is calculated as the mean squared error (MSE) between this estimate and provided approximations of clean measurements used as label values in the case where true parameter estimates are not available. Thus, it is desirable that these labels contain as little noise as possible, although estimation can still take place with noisy labels, as will be seen in Sec.~\ref{sec_results}. Clean label values are realized by averaging over many input groups sharing the same parameter values and using this same average for all of the contributing input trajectory groups during training. See Sec.~\ref{sec_data} for more details. In this way the model can be viewed as a denoising autoencoder \cite{goodfellow}, with de-noised output provided for every noisy input example, and with a latent space corresponding to the physical parameters being estimated.

In our model, the parameters learned during training are the various weights and biases of the neural networks in the encoder and the decoder. Notably, we are not directly estimating master equation parameters during training, but rather learning a function that maps noisy measurement records to physical parameters. This approach differs from that of Ref.~\cite{PRXQuantum.2.040355}, where the physical parameters are the result of an optimization and no such function is learned. One advantage of our approach is that it enables extremely fast parameter estimation once the model is trained, even when observing systems with parameter sets not included in the training data. Furthermore, in the case where the training data includes true physical parameter values, it allows us to learn the map with the encoder only, bypassing the need for a physical model entirely.
In this case, scaling is limited by the amount and type of training data required, rather than system size. The amount of training data required will vary by application and requires further investigation.
Finally, learning a map for physical parameters as a function of measurement records allows for the possibility of fewer measurements being required for prediction versus training as the model learns to account for noise in the input data, as illustrated by results in \ref{sec_results} showing how estimation accuracy improves with training set size even when the number of input records is held constant.

A second difference between our approach and that of Ref.~\cite{PRXQuantum.2.040355} is that when labeled training data is unavailable and a physical decoder is necessary, the flexible correction scheme in the decoder allows the model to compensate for non-Markovian or nonlinear dynamics since it is not bound by a Lindblad form, while still being informed by the master equation for a first order model of the dynamics.

\subsection{Training and Data}
\label{sec_data}
In order to generate the data for training, validation, and testing, we use an Euler-Maruyama integrator to solve Eq.~\eqref{eqn:volt} for both qubits for each of $40$ values of $\epsilon$ evenly spaced on $[0,2)$ with fixed $\Omega = 1.395$ radians/$\mu$s, and the same number of $\Omega$ values evenly spaced on $[1,5)$ with fixed $\epsilon = 1.0$ radians/$\mu$s for a total of $K = 80$ ($\Omega$, $\epsilon$) pairs. For each pair of values ($\Omega_k$, $\epsilon_k$), $N = 32{,}000$ measurement trajectories are simulated for $T = 4 \mu s$. Half of the collection of pair of values ($\Omega_k$, $\epsilon_k$), corresponding to every other parameter pair, is used for training. The other half, corresponding to parameters midway between training values, is split evenly in the number of trajectories to be used for validation and testing. This ensures the training set contains a disjoint set of parameter pairs from the validation/test set. The two endpoints of the validation/test set are excluded from each end to ensure that the training data extends slightly beyond the domain of validation/test values. Only weak measurement records are being used, and no strong measurements are being performed. 

Values of $\kappa = 3.326$ radians/$\mu$s and $\eta = 0.1469$ are selected to be consistent with the example of superconducting qubits as found, for example, in Ref.~\cite{PRXQuantum.2.040355}, though with a stronger measurement back-action $\kappa$ after simulations showed a larger value to offer a good balance between dephasing rate and measurement noise. While either $\Omega$, $\epsilon$, or both are treated as unknown parameters and comprise the latent space of our model, it is assumed in our simulations that $\kappa$ and $\eta$ are known with high precision as they have already been calibrated. Our approach can be used to perform this calibration step, since the latent space of the model is not limited to Hamiltonian parameters and could support any combination of Hamiltonian and Lindblad parameters as unknown variables. 
Alternatively, other calibration methods found in the literature can be used to estimate them, such as in \cite{eta_est}. It should be noted, however, that in practice if calibration errors are present in the model values, the degrees of freedom in the decoder should make parameter estimation robust to these errors as demonstrated in what is likely the more challenging case of unanticipated single particle relaxation examined in Sec.~\ref{sec_decoder}. The impact on parameter estimation accuracy is expected to be small for small calibration errors. A study quantifying this dependency is an interesting area for future work.

During training, for each true parameter set $\theta_k = (\Omega_k, \epsilon_k)$, trajectory groups of a preset size $d$ are randomly selected from the full training set and their averages are provided as input values to the model, such that each mini-batch is comprised of $M = N/d$ averaged trajectories $\{\tilde{x}_{j,k}(t)\}_{j=1}^M$ where $\tilde{x}_{j,k}(t) = \frac{1}{d}\sum_{i\in I} r_{i,k}(t)$, $r_{i,k}(t)$ is measurement record $i$ for parameter set $k$, and $I$ is a set of trajectory indices of size $d$ randomly selected from $1,2,...,N$ without replacement until all $N$ trajectories are used, which defines one epoch of training. In this way, each mini-batch consists of a different set of noisy trajectories as input to maximize the diversity of training examples. In the unsupervised case, for all $j=1,...,M$, the value $x_k(t)$ to be used in the loss function is the average over the full training set of trajectories associated with the true parameter set $\theta_k$, such that $x_k(t) = \frac{1}{N}\sum_{i=1}^N r_{i,k}(t)$. 

To make the interpretation of our model as a denoising autoencoder more concrete, we note that DNAs are characterized by a corruption process $C(\tilde{y}\vert y)$ whereby noisy inputs $\tilde{y}$ are generated for each uncorrupted training example $y$ \cite{goodfellow}. In our case, the full trajectory means $x_k(t)$ take on the role of $y$, approximating the solution to the unconditioned SME, while the random selection of much smaller trajectory groups is the process by which the corrupted data elements $\tilde{y}$ are generated.

In the supervised case, the loss function is the MSE between predicted parameter sets $\tilde{\theta}_k$ and the provided true values $\theta_k$
\begin{equation}
    \mathcal{L} = \frac{1}{pK}\sum_k \left\vert\left\vert \tilde{\theta}_k - \theta_k \right\vert\right\vert_2^2,
    \label{eqn:sup_loss}
\end{equation}
where $p$ is the dimension of $\theta_k$,
the number of parameters being estimated. This is further averaged over the $M$ groups of size $d$ when multiple trajectory groups are provided as input. In the unsupervised case, the loss function is the MSE between each $x_k$ and the estimated clean measurement record 
\begin{equation}
    \mathcal{L} = \frac{1}{MKN_t}\sum_{j,k,t}\left\vert x_k(t) - \mathcal{M}(\tilde{x}_{j,k})(t)\right\vert^2
    \label{eqn:unsup_loss}
\end{equation} where $\mathcal{M}$ denotes the model, $t$ is the time index, and $N_t \equiv T/\Delta t$ is the number of time points excluding the initial condition.

During evaluation, validation and test error is evaluated in a similar manner to how training mini-batches are selected, with groups of size $d$ randomly chosen from the full validation and test sets, followed by the calculation of the MSE. This process is repeated 100 times and the average is taken as the calculated MSE for each set. Note that for some scenarios only one group will be available as input, either because the data is noise free or because the group size equals the dataset size, in which case the shuffling has no effect.

Training is performed for multiple runs of 100 epochs, with a learning rate of $3\times 10^{-3}$ and a decay rate of 0.99 per epoch, resetting the learning rate after each run, continuing until the validation loss drops by less than five percent run-to-run, and for the last 20 epochs of the final run. Once converged, parameter MSE is computed for the validation and test sets. This entire process is performed 100 times, each with a different random initialization of the model parameters. The best model is considered to be the one with the smallest validation loss after the final epoch, and this model is used to evaluate the MSE for the test set. Hyperparameter tuning for the model layer sizes was performed using a grid search for a physical system with parameters distinct from those in Sec.~\ref{sec_results}.
\section{Evaluation}
\label{sec_results}
To fully assess the performance of the model, we evaluate two distinct cases: the encoder alone to evaluate the case of labeled training data, and the full model for the case where labels are not available. We consider a range of trajectory group sizes $d$ as well as noise-free data derived from an Euler integration of the unconditioned master equation. The noise-free case is denoted by $\infty$ as the number of trajectories in all tables.

To minimize the impact of the weak measurement back-action via the parameter $\kappa$, in Secs.~\ref{sec_encoder} and \ref{sec_full_model} the initial state is chosen to be spin-up in the directions of measurement, and the directions of $X$ and $Y$ were used for the first and second qubit, respectively. A different configuration is used in Sec.~\ref{sec_decoder}, where more diverse measurements are needed to correct for unanticipated single particle relaxation not explicitly present in the decoder's physical model.

\subsection{Supervised Learning}
\label{sec_encoder}

First, we study the impact of noise in the training set and how it helps or hinders the model's performance when predicting parameters from noisy data. We do this by performing estimation with two models, one trained using noise-free measurement values, and another set trained using eight groups of $d = 4{,}000$ averaged measurement records. Both models are then evaluated on a test set of noisy, averaged measurement records with the same group size $d$ as the noisy training set. In both cases, the model with the best validation loss on its respective validation set out of 100 randomly initialized models was used.

\begin{figure}[th!]
    \centering
    \includegraphics[width=0.4\textwidth]{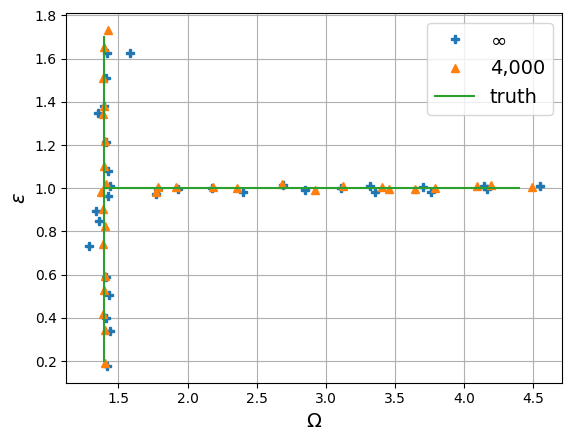}
    \includegraphics[width=0.41\textwidth]{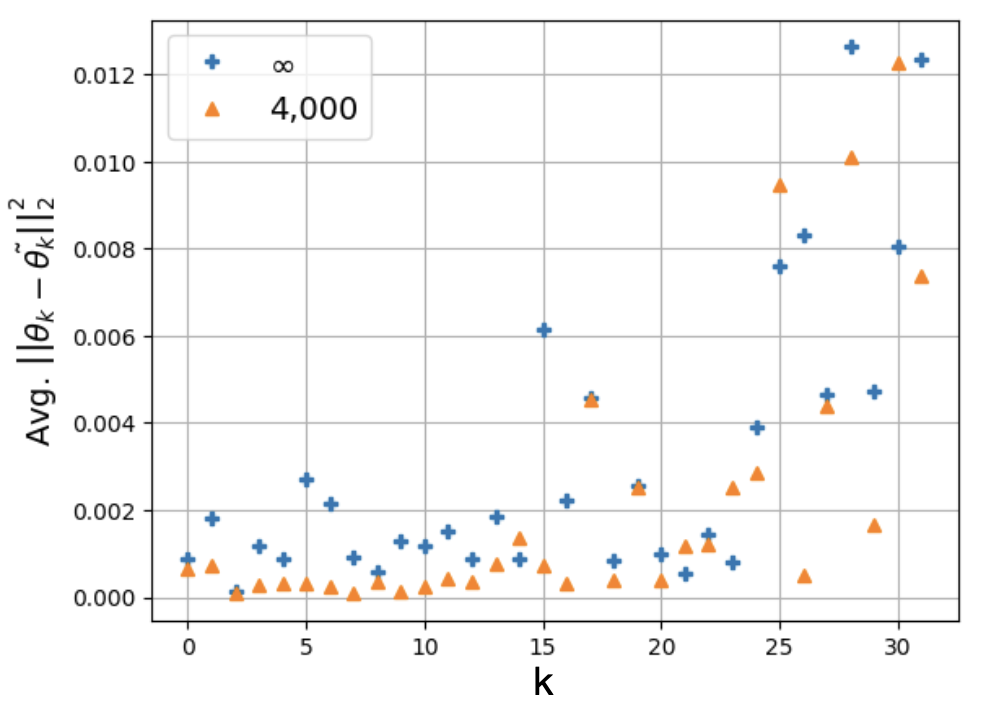}
    \caption{Estimations of the parameter pair $\theta_k = (\Omega_k, \epsilon_k)$ (top) and the squared error (bottom) averaged over four noisy trajectory groups of $d = 4{,}000$ measurement records for a model trained on clean data and noisy data, then evaluated on noisy data. Shuffle evaluated test set MSEs are (8.67e-3, 5.83e-4) and (4.94e-3, 4.87e-4) respectively.}
    \label{fig:encoder}
\end{figure}

Figure \ref{fig:encoder} shows an example of the error of the estimated parameter pair $\tilde{\theta}_k$ compared to truth for each of the 32 $k$ values in the trimmed test set. We see that for many specific pairs and for the overall MSE, the model trained on the noise-free measurements is producing less accurate estimates than the model trained on noisy data. This is consistent in both the $\Omega$ and $\epsilon$ errors. This suggests that the model learns to account for noise, as expected, and training with noise on the level of measurements used for prediction is beneficial.

Next we examine the impact of the training group size $d$. Table \ref{tbl:mse_encoder_only} lists test set MSEs for a number of training sets where evaluation is taking place with the same group size $d$ used for training. The best and median were taken from the list of results sorted according to validation loss, as this is what would be known in practice during training. Here we see a steady improvement as the number of trajectories in a group increases, and noise decreases, saturating in the case where noise-free data is used for training and evaluation. We note that this test differs from the results in Fig.~\ref{fig:encoder} where the model trained without noise was evaluated on noisy data, as here we consider only the case where training and test group sizes are equal.

\begin{table}[!hbt]
\begin{tabular}{|c|c|c|c|}
\hline
$d$ & Best MSE & Median MSE & Mean MSE \\ \hline
2{,}000    & 9.87e-3, 8.76e-4 & 9.72e-3, 8.71e-4 & 9.75e-3, 9.79e-4 \\ \hline
4{,}000    & 4.94e-3, 4.87e-4 & 5.23e-3, 5.44e-4 & 5.62e-3, 5.75e-4 \\ \hline
8{,}000    & 4.17e-3, 2.74e-4 & 2.94e-3, 2.74e-4 & 3.38e-3, 4.13e-4 \\ \hline
16{,}000    & 2.19e-3, 2.06e-4 & 1.40e-3, 1.92e-4 & 1.80e-3, 2.94e-4 \\ \hline
$\infty$    & 1.01e-5, 2.04e-5 & 4.51e-6, 4.82e-5 & 1.32e-5, 5.68e-5 \\ \hline
\end{tabular}
\caption{Each cell shows the MSE pair for ($\Omega$, $\epsilon$) estimates on supervised training set using the encoder only.}
\label{tbl:mse_encoder_only}
\end{table}

\subsection{Unsupervised Learning}
\label{sec_full_model}

We now consider the unsupervised case where the labels of the training data are not known, but it is assumed that the physical model in the decoder is correct, i.e.\ we are still not learning parameters for the drift function correction $f_{\text{LSTM}}$, which we will consider in Sec.~\ref{sec_decoder}. We are therefore using the MSE in Eq.~\eqref{eqn:unsup_loss} as the loss function. 

First we examine how parameter estimatation accuracy for a fixed group size $d = 4{,}000$ varies with the total training set size $N$. Table \ref{tbl:train_size} shows $(\Omega, \epsilon)$ test set MSEs for the case where $\Omega$ is unknown but fixed at a true value of 1.395 radians/$\mu$s and $\epsilon$ is allowed to vary, for various training set sizes $N$. The same test set containing $16{,}000$ trajectories is used for each row. From the table, we see that accuracy improves significantly as the amount of training data increases, even though the groups being presented to the model for parameter estimation remain the same. This indicates that a greater diversity of noisy measurement records when training results in models that can produce more accurate parameter estimates when presented with the same number of measurements when performing prediction. This motivates using $N = 32{,}000$ going forward.

\begin{table}[!hbt]
\centering
\begin{tabular}{|c|c|c|c|}
\hline
$N$ & Best MSE & Median MSE & Mean MSE \\ \hline
4{,}000    & 1.51e-3, 1.12e-3 & 6.58e-4, 9.86e-4 & 8.33e-4, 1.08e-3 \\ \hline
8{,}000    & 3.27e-4, 7.71e-4 & 5.20e-4, 7.94e-4 & 5.38e-4, 8.81e-4 \\ \hline
16{,}000    & 1.83e-4, 7.49e-4 & 3.20e-4, 8.41e-4 & 3.07e-4, 8.38e-4 \\ \hline
32{,}000    & 8.14e-5, 7.16e-4 & 9.82e-5, 8.15e-4 & 1.45e-4, 8.12e-4 \\ \hline
\end{tabular}
\caption{MSE of $(\Omega, \epsilon)$ when using $d = 4{,}000$ trajectories to estimate parameters for varying training set sizes $N$ for an unsupervised training set with fixed $\Omega$ and varying $\epsilon$. The values of $\kappa$ and $\eta$ are assumed to be known exactly.}
\label{tbl:train_size}
\end{table}

Next, we consider the impact of group size $d$ and measurement record time spacing $\Delta t$ on accuracy. Table \ref{tbl:mse_single_omega_shuffle} again shows $(\Omega, \epsilon)$ test set MSEs for the case where $\Omega$ is unknown but fixed at a true value of 1.395 radians/$\mu$s and $\epsilon$ is allowed to vary.  It illustrates how extremely low test set MSEs are achievable for $\Omega$ in this case, which is expected given the high volume of training data available for a single value. The accuracy of $\epsilon$ estimates depends on both the number of trajectories used to create each input sequence, as well as the time spacing at which measurements are recorded. To evaluate measurement records with varying time spacing, trajectories simulated with $\Delta t = 2^{-8} \mu$s are sub-sampled to avoid introducing numerical integration error associated with simulations with a larger time step.

Table \ref{tbl:mse_omega_eps_shuffle} shows test set MSEs for the training set where both $\Omega$ and $\epsilon$ are allowed to vary. This is a harder task as the model has less training data for each unique $\Omega$ value, hence the loss in accuracy for that parameter. The error in $\epsilon$ remains roughly the same or better as in the fixed $\Omega$ case, with a best case root mean-square error of around one percent of the median test value of the $\epsilon$ parameter. Here we see a roughly linear trend in the MSE versus $d$ at smaller time steps, as doubling the number of trajectories in a group roughly halves the MSE. This trend breaks down, however, for the $\Delta t = 2^{-4} \mu$s case, suggesting a permissive time step threshold around $\Delta t = 2^{-6} \mu$s at or below which the expected trend in accuracy versus input data size is realized.

\begin{table}[]
\begin{tabular}{|c|c|c|c|}
\hline
$d$ & $\Delta t = 2^{-8} \,\mu s$ & $2^{-6}$ & $2^{-4}$ \\ \hline
2{,}000    & 5.78e-5, 1.40e-3 & 1.20e-4, 1.40e-3 & 1.22e-4, 2.83e-3 \\ \hline
4{,}000    & 8.14e-5, 7.16e-4 & 1.66e-4, 7.90e-4 & 1.19e-4, 2.09e-3 \\ \hline
8{,}000    & 8.51e-5, 3.42e-4 & 7.16e-5, 3.63e-4 & 8.02e-5, 1.82e-3 \\ \hline
16{,}000    & 8.31e-5, 1.57e-4 & 1.66e-4, 1.57e-4 & 1.19e-4, 1.89e-3 \\ \hline
$\infty$    & 1.66e-6, 6.08e-6 & 7.21e-6, 1.04e-4 & 1.52e-5, 1.71e-3 \\ \hline
\end{tabular}
\caption{MSE of ($\Omega$, $\epsilon$) estimates on unsupervised training set with fixed $\Omega$ and varying $\epsilon$ and $N = 32,000$. The values of $\kappa$ and $\eta$ are assumed to be known exactly.}
\label{tbl:mse_single_omega_shuffle}
\end{table}

\begin{table}[]
\begin{tabular}{|c|c|c|c|}
\hline
$d$ & $\Delta t = 2^{-8} \,\mu s$ & $2^{-6}$ & $2^{-4}$ \\ \hline
2{,}000    & 1.12e-2, 8.01e-4 & 1.10e-2, 8.42e-4 & 1.23e-2, 2.12e-3 \\ \hline
4{,}000    & 5.87e-3, 4.56e-4 & 6.46e-3, 5.07e-4 & 6.39e-3, 1.68e-3 \\ \hline
8{,}000    & 3.08e-3, 2.32e-4 & 2.98e-3, 2.87e-4 & 3.34e-3, 1.46e-3 \\ \hline
16{,}000    & 1.68e-3, 1.19e-4 & 1.11e-3, 1.63e-4 & 1.58e-3, 1.39e-3 \\ \hline
$\infty$  & 1.25e-5, 7.75e-6 & 1.76e-5, 7.07e-5 & 1.14e-4, 1.23e-3 \\ \hline
\end{tabular}
\caption{MSE of ($\Omega$, $\epsilon$) estimates on unsupervised training set with varying $\Omega$ and $\epsilon$ and $N = 32,000$. The values of $\kappa$ and $\eta$ are assumed to be known exactly.}
\label{tbl:mse_omega_eps_shuffle}
\end{table}

\subsection{Model Correction}
\label{sec_decoder}

In this section we demonstrate the ability of the decoder to correct for dynamics not explicitly considered in the physical model (Eq.~\eqref{eqn:sme}). This is done by enabling training for the decoder LSTM parameters. For the datasets, we simulate 30{,}000 trajectories with fixed $\Omega = 1.395$ radians/$\mu$s that is known to the model, but with varying unknown $\epsilon$ which is the parameter to be estimated. We add a dissipative term for each qubit corresponding to the Lindblad operator $\mathcal{D}[\sqrt{\gamma_s}\sigma_i^-](\rho)$ in Eq.~\eqref{eqn:sme}, where $\gamma_s = 0.1$ and $\sigma_i^- = \frac{1}{2}(X_i - iY_i)$ is the single particle relaxation operator mapping the excited state to the ground state. Note that while the choice of a single unknown parameter is made in this section to provide a clear and straightforward demonstration, the model still supports multiple possible parameters to be estimated in addition to the decoder degrees of freedom.

In this case, we take a more diverse set of measurements, simulating 10{,}000 trajectories measuring in each of the $X$, $Y$, and $Z$ directions for both qubits, and relaxing $\kappa$ to one-fourth the value used in the last section to reduce measurement back-action. Spin-up in the $Z$ direction is the initial state for each qubit. This more cautious approach to measurement is warranted if completely unknown effects are expected to be present. More information about the type of phenomenon, but not necessarily the magnitude, could allow for a more targeted measurement scheme, but here we keep it general. The trajectory group size used for each input was $d = 5{,}000$, and the best of 20 randomly initialized models was selected for the results in this section.

\begin{figure}[!t]
    \centering
    \includegraphics[width=0.4\textwidth]{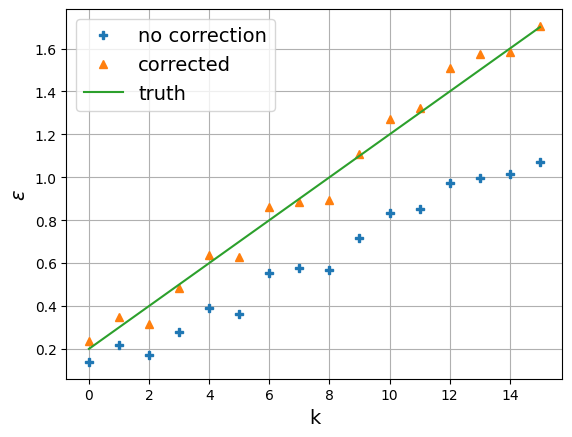}
    \includegraphics[width=0.4\textwidth]{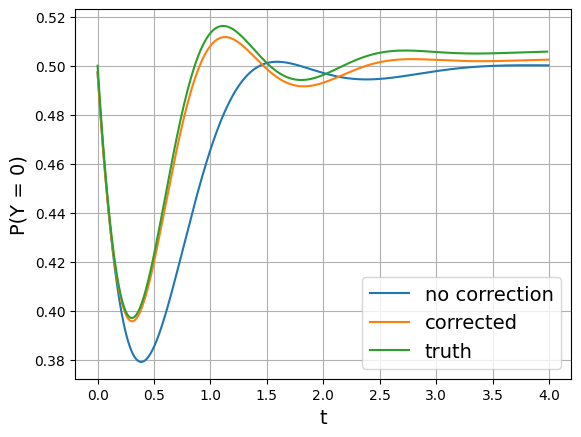}
    \caption{Estimate of $\epsilon$ (top) and $P(Y_i=0)$ when $\epsilon = 1.7$ (bottom) and measuring $X$ with spin-up in $Z$ as the initial state for both qubits for training data simulated with single particle relaxation rate $\gamma_s = 0.1$ but a model that does not explicitly account for $\gamma_s$. Results are shown for the physical model alone and with a correction learned by the decoder LSTM.}
    \label{fig:decoder}
\end{figure}

The results of the parameter estimation with and without the correction are shown in Fig.~\ref{fig:decoder} and Table~ \ref{tbl:mse_xyz_shuffle}. Here we see that single particle relaxation has introduced a significant bias to the estimated $\epsilon$ parameters when unaccounted for in the model, but the decoder LSTM has successfully corrected for the effect, returning the MSE to a value much closer to where it would have been had the physical model explicitly accounted for it.

\begin{table}[hbt]
\begin{tabular}{|c|c|c|c|}
\hline
$d$ & $\gamma_s = 0.1$ & $\gamma_s = 0.0$ &  $\gamma_s = 0.0$ + correction \\ \hline
5{,}000    & 2.84e-3 & 0.142 & 3.59e-3 \\ \hline 
$\infty$  & 2.12e-6 & 0.145 & 3.26e-4 \\ \hline
\end{tabular}
\caption{MSE of $\epsilon$ estimates using different $\gamma_s$ values in the model. The final column shows results when learning is enabled for the free parameters in the decoder to account for the unanticipated term $\gamma_s$. Measurement time spacing is $\Delta t = 2^{-8}\mu s$.}
\label{tbl:mse_xyz_shuffle}
\end{table}

\section{Conclusions} \label{sec:conclusions}

We have proposed a machine learning model based on a denoising autoencoder capable of direct estimation of physical parameters in a system modeled by a stochastic master equation from weak measurement records. The model is capable of learning in a supervised and an unsupervised context, and it can accurately predict parameters for systems not seen in the training data. While leveraging the use of a master equation integrator to enable unsupervised learning, the autoencoder is robust to unanticipated dynamics not included in its physical model. We have demonstrated this in the case of unanticipated Lindblad dissipative terms.
Previous work has established that LSTM models are capable of learning dynamics beyond unknown Lindbladian dissipation \cite{siddiqi_rnn2,nm_rnn}, and it is an interesting subject for future work to examine the robustness of our model to the presence of unanticipated non-Markovian and nonlinear dynamics in the training data.

Another potential subject for future investigation is the ability of the model to estimate parameters for much larger systems in the supervised context, where it may not necessarily be subject to the exponential scaling of the Hilbert space with system size that often limits other approaches to parameter estimation.
\begin{acknowledgments}
This material is based upon work supported by the National Science Foundation the Quantum Leap Big Idea under Grant No. OMA-1936388. AKR and CM acknowledge support from NSF QLCI OMA-2016244.

This work was performed, in part, at the Center for Integrated Nanotechnologies, an Office of Science User Facility operated for the U.S. Department of Energy (DOE) Office of Science.

This article has been co-authored by an employee of National Technology \& Engineering Solutions of Sandia, LLC under Contract No. DE-NA0003525 with the U.S. Department of Energy (DOE). The employee owns all right, title and interest in and to the article and is solely responsible for its contents. The United States Government retains and the publisher, by accepting the article for publication, acknowledges that the United States Government retains a non-exclusive, paid-up, irrevocable, world-wide license to publish or reproduce the published form of this article or allow others to do so, for United States Government purposes. The DOE will provide public access to these results of federally sponsored research in accordance with the DOE Public Access Plan \url{https://www.energy.gov/downloads/doe-public-access-plan.}
\end{acknowledgments}
%

\begin{thebibliography}{61}%
\makeatletter
\providecommand \@ifxundefined [1]{%
 \@ifx{#1\undefined}
}%
\providecommand \@ifnum [1]{%
 \ifnum #1\expandafter \@firstoftwo
 \else \expandafter \@secondoftwo
 \fi
}%
\providecommand \@ifx [1]{%
 \ifx #1\expandafter \@firstoftwo
 \else \expandafter \@secondoftwo
 \fi
}%
\providecommand \natexlab [1]{#1}%
\providecommand \enquote  [1]{``#1''}%
\providecommand \bibnamefont  [1]{#1}%
\providecommand \bibfnamefont [1]{#1}%
\providecommand \citenamefont [1]{#1}%
\providecommand \href@noop [0]{\@secondoftwo}%
\providecommand \href [0]{\begingroup \@sanitize@url \@href}%
\providecommand \@href[1]{\@@startlink{#1}\@@href}%
\providecommand \@@href[1]{\endgroup#1\@@endlink}%
\providecommand \@sanitize@url [0]{\catcode `\\12\catcode `\$12\catcode
  `\&12\catcode `\#12\catcode `\^12\catcode `\_12\catcode `\%12\relax}%
\providecommand \@@startlink[1]{}%
\providecommand \@@endlink[0]{}%
\providecommand \url  [0]{\begingroup\@sanitize@url \@url }%
\providecommand \@url [1]{\endgroup\@href {#1}{\urlprefix }}%
\providecommand \urlprefix  [0]{URL }%
\providecommand \Eprint [0]{\href }%
\providecommand \doibase [0]{https://doi.org/}%
\providecommand \selectlanguage [0]{\@gobble}%
\providecommand \bibinfo  [0]{\@secondoftwo}%
\providecommand \bibfield  [0]{\@secondoftwo}%
\providecommand \translation [1]{[#1]}%
\providecommand \BibitemOpen [0]{}%
\providecommand \bibitemStop [0]{}%
\providecommand \bibitemNoStop [0]{.\EOS\space}%
\providecommand \EOS [0]{\spacefactor3000\relax}%
\providecommand \BibitemShut  [1]{\csname bibitem#1\endcsname}%
\let\auto@bib@innerbib\@empty
\bibitem [{\citenamefont {Preskill}(2018)}]{preskill_nisq}%
  \BibitemOpen
  \bibfield  {author} {\bibinfo {author} {\bibfnamefont {J.}~\bibnamefont
  {Preskill}},\ }\bibfield  {title} {\bibinfo {title} {Quantum {C}omputing in
  the {NISQ} era and beyond},\ }\href
  {https://doi.org/10.22331/q-2018-08-06-79} {\bibfield  {journal} {\bibinfo
  {journal} {{Quantum}}\ }\textbf {\bibinfo {volume} {2}},\ \bibinfo {pages}
  {79} (\bibinfo {year} {2018})}\BibitemShut {NoStop}%
\bibitem [{\citenamefont {Sarovar}\ \emph {et~al.}(2020)\citenamefont
  {Sarovar}, \citenamefont {Proctor}, \citenamefont {Rudinger}, \citenamefont
  {Young}, \citenamefont {Nielsen},\ and\ \citenamefont
  {Blume-Kohout}}]{crosstalk_sarovar}%
  \BibitemOpen
  \bibfield  {author} {\bibinfo {author} {\bibfnamefont {M.}~\bibnamefont
  {Sarovar}}, \bibinfo {author} {\bibfnamefont {T.}~\bibnamefont {Proctor}},
  \bibinfo {author} {\bibfnamefont {K.}~\bibnamefont {Rudinger}}, \bibinfo
  {author} {\bibfnamefont {K.}~\bibnamefont {Young}}, \bibinfo {author}
  {\bibfnamefont {E.}~\bibnamefont {Nielsen}},\ and\ \bibinfo {author}
  {\bibfnamefont {R.}~\bibnamefont {Blume-Kohout}},\ }\bibfield  {title}
  {\bibinfo {title} {Detecting crosstalk errors in quantum information
  processors},\ }\href {https://doi.org/10.22331/q-2020-09-11-321} {\bibfield
  {journal} {\bibinfo  {journal} {{Quantum}}\ }\textbf {\bibinfo {volume}
  {4}},\ \bibinfo {pages} {321} (\bibinfo {year} {2020})}\BibitemShut {NoStop}%
\bibitem [{\citenamefont {Tripathi}\ \emph {et~al.}(2022)\citenamefont
  {Tripathi}, \citenamefont {Chen}, \citenamefont {Khezri}, \citenamefont
  {Yip}, \citenamefont {Levenson-Falk},\ and\ \citenamefont
  {Lidar}}]{crosstalk_lidar}%
  \BibitemOpen
  \bibfield  {author} {\bibinfo {author} {\bibfnamefont {V.}~\bibnamefont
  {Tripathi}}, \bibinfo {author} {\bibfnamefont {H.}~\bibnamefont {Chen}},
  \bibinfo {author} {\bibfnamefont {M.}~\bibnamefont {Khezri}}, \bibinfo
  {author} {\bibfnamefont {K.-W.}\ \bibnamefont {Yip}}, \bibinfo {author}
  {\bibfnamefont {E.}~\bibnamefont {Levenson-Falk}},\ and\ \bibinfo {author}
  {\bibfnamefont {D.~A.}\ \bibnamefont {Lidar}},\ }\bibfield  {title} {\bibinfo
  {title} {Suppression of crosstalk in superconducting qubits using dynamical
  decoupling},\ }\href {https://doi.org/10.1103/PhysRevApplied.18.024068}
  {\bibfield  {journal} {\bibinfo  {journal} {Phys. Rev. Appl.}\ }\textbf
  {\bibinfo {volume} {18}},\ \bibinfo {pages} {024068} (\bibinfo {year}
  {2022})}\BibitemShut {NoStop}%
\bibitem [{\citenamefont {Kitaev}(1997)}]{Kitaev1997}%
  \BibitemOpen
  \bibfield  {author} {\bibinfo {author} {\bibfnamefont {A.~Y.}\ \bibnamefont
  {Kitaev}},\ }\bibfield  {title} {\bibinfo {title} {Quantum computations:
  algorithms and error correction},\ }\href
  {https://doi.org/10.1070/RM1997v052n06ABEH002155} {\bibfield  {journal}
  {\bibinfo  {journal} {Russian Mathematical Surveys}\ }\textbf {\bibinfo
  {volume} {52}},\ \bibinfo {pages} {1191} (\bibinfo {year}
  {1997})}\BibitemShut {NoStop}%
\bibitem [{\citenamefont {Aharonov}\ and\ \citenamefont
  {Ben-Or}(1997)}]{Aharonov1997}%
  \BibitemOpen
  \bibfield  {author} {\bibinfo {author} {\bibfnamefont {D.}~\bibnamefont
  {Aharonov}}\ and\ \bibinfo {author} {\bibfnamefont {M.}~\bibnamefont
  {Ben-Or}},\ }\bibfield  {title} {\bibinfo {title} {Fault-tolerant quantum
  computation with constant error},\ }in\ \href
  {https://doi.org/10.1145/258533.258579} {\emph {\bibinfo {booktitle}
  {Proceedings of the Twenty-Ninth Annual ACM Symposium on Theory of
  Computing}}},\ \bibinfo {series and number} {STOC '97}\ (\bibinfo
  {publisher} {Association for Computing Machinery},\ \bibinfo {address} {New
  York, NY, USA},\ \bibinfo {year} {1997})\ p.\ \bibinfo {pages}
  {176–188}\BibitemShut {NoStop}%
\bibitem [{\citenamefont {Preskill}(1998)}]{Preskill1998}%
  \BibitemOpen
  \bibfield  {author} {\bibinfo {author} {\bibfnamefont {J.}~\bibnamefont
  {Preskill}},\ }\bibfield  {title} {\bibinfo {title} {Reliable quantum
  computers},\ }\href {https://doi.org/10.1098/rspa.1998.0167} {\bibfield
  {journal} {\bibinfo  {journal} {Proceedings of the Royal Society of London.
  Series A: Mathematical, Physical and Engineering Sciences}\ }\textbf
  {\bibinfo {volume} {454}},\ \bibinfo {pages} {385} (\bibinfo {year}
  {1998})}\BibitemShut {NoStop}%
\bibitem [{\citenamefont {Knill}\ \emph {et~al.}(1998)\citenamefont {Knill},
  \citenamefont {Laflamme},\ and\ \citenamefont {Zurek}}]{Knill1998}%
  \BibitemOpen
  \bibfield  {author} {\bibinfo {author} {\bibfnamefont {E.}~\bibnamefont
  {Knill}}, \bibinfo {author} {\bibfnamefont {R.}~\bibnamefont {Laflamme}},\
  and\ \bibinfo {author} {\bibfnamefont {W.~H.}\ \bibnamefont {Zurek}},\
  }\bibfield  {title} {\bibinfo {title} {Resilient quantum computation},\
  }\href {https://doi.org/10.1126/science.279.5349.342} {\bibfield  {journal}
  {\bibinfo  {journal} {Science}\ }\textbf {\bibinfo {volume} {279}},\ \bibinfo
  {pages} {342} (\bibinfo {year} {1998})}\BibitemShut {NoStop}%
\bibitem [{\citenamefont {Gottesman}(1998)}]{Gottesman1998}%
  \BibitemOpen
  \bibfield  {author} {\bibinfo {author} {\bibfnamefont {D.}~\bibnamefont
  {Gottesman}},\ }\bibfield  {title} {\bibinfo {title} {Theory of
  fault-tolerant quantum computation},\ }\href
  {https://doi.org/10.1103/PhysRevA.57.127} {\bibfield  {journal} {\bibinfo
  {journal} {Phys. Rev. A}\ }\textbf {\bibinfo {volume} {57}},\ \bibinfo
  {pages} {127} (\bibinfo {year} {1998})}\BibitemShut {NoStop}%
\bibitem [{\citenamefont {Carleo}\ and\ \citenamefont
  {Troyer}(2017)}]{carleo_troyer}%
  \BibitemOpen
  \bibfield  {author} {\bibinfo {author} {\bibfnamefont {G.}~\bibnamefont
  {Carleo}}\ and\ \bibinfo {author} {\bibfnamefont {M.}~\bibnamefont
  {Troyer}},\ }\bibfield  {title} {\bibinfo {title} {Solving the quantum
  many-body problem with artificial neural networks},\ }\href
  {https://doi.org/10.1126/science.aag2302} {\bibfield  {journal} {\bibinfo
  {journal} {Science}\ }\textbf {\bibinfo {volume} {355}},\ \bibinfo {pages}
  {602} (\bibinfo {year} {2017})}\BibitemShut {NoStop}%
\bibitem [{\citenamefont {Carleo}\ \emph {et~al.}(2018)\citenamefont {Carleo},
  \citenamefont {Nomura},\ and\ \citenamefont {Imada}}]{carleo_2018}%
  \BibitemOpen
  \bibfield  {author} {\bibinfo {author} {\bibfnamefont {G.}~\bibnamefont
  {Carleo}}, \bibinfo {author} {\bibfnamefont {Y.}~\bibnamefont {Nomura}},\
  and\ \bibinfo {author} {\bibfnamefont {M.}~\bibnamefont {Imada}},\ }\bibfield
   {title} {\bibinfo {title} {Constructing exact representations of quantum
  many-body systems with deep neural networks},\ }\href
  {https://doi.org/10.1038/s41467-018-07520-3} {\bibfield  {journal} {\bibinfo
  {journal} {Nature Communications}\ }\textbf {\bibinfo {volume} {9}},\
  \bibinfo {pages} {5322} (\bibinfo {year} {2018})}\BibitemShut {NoStop}%
\bibitem [{\citenamefont {Carrasquilla}\ \emph {et~al.}(2019)\citenamefont
  {Carrasquilla}, \citenamefont {Torlai}, \citenamefont {Melko},\ and\
  \citenamefont {Aolita}}]{qst_1810_10584}%
  \BibitemOpen
  \bibfield  {author} {\bibinfo {author} {\bibfnamefont {J.}~\bibnamefont
  {Carrasquilla}}, \bibinfo {author} {\bibfnamefont {G.}~\bibnamefont
  {Torlai}}, \bibinfo {author} {\bibfnamefont {R.~G.}\ \bibnamefont {Melko}},\
  and\ \bibinfo {author} {\bibfnamefont {L.}~\bibnamefont {Aolita}},\
  }\bibfield  {title} {\bibinfo {title} {Reconstructing quantum states with
  generative models},\ }\href {https://doi.org/10.1038/s42256-019-0028-1}
  {\bibfield  {journal} {\bibinfo  {journal} {Nature Machine Intelligence}\
  }\textbf {\bibinfo {volume} {1}},\ \bibinfo {pages} {155} (\bibinfo {year}
  {2019})}\BibitemShut {NoStop}%
\bibitem [{\citenamefont {Rocchetto}\ \emph {et~al.}(2018)\citenamefont
  {Rocchetto}, \citenamefont {Grant}, \citenamefont {Strelchuk}, \citenamefont
  {Carleo},\ and\ \citenamefont {Severini}}]{qst_vae}%
  \BibitemOpen
  \bibfield  {author} {\bibinfo {author} {\bibfnamefont {A.}~\bibnamefont
  {Rocchetto}}, \bibinfo {author} {\bibfnamefont {E.}~\bibnamefont {Grant}},
  \bibinfo {author} {\bibfnamefont {S.}~\bibnamefont {Strelchuk}}, \bibinfo
  {author} {\bibfnamefont {G.}~\bibnamefont {Carleo}},\ and\ \bibinfo {author}
  {\bibfnamefont {S.}~\bibnamefont {Severini}},\ }\bibfield  {title} {\bibinfo
  {title} {Learning hard quantum distributions with variational autoencoders},\
  }\href {https://doi.org/10.1038/s41534-018-0077-z} {\bibfield  {journal}
  {\bibinfo  {journal} {npj Quantum Information}\ }\textbf {\bibinfo {volume}
  {4}},\ \bibinfo {pages} {28} (\bibinfo {year} {2018})}\BibitemShut {NoStop}%
\bibitem [{\citenamefont {Carrasquilla}\ and\ \citenamefont
  {Torlai}(2021)}]{qst_tutorial}%
  \BibitemOpen
  \bibfield  {author} {\bibinfo {author} {\bibfnamefont {J.}~\bibnamefont
  {Carrasquilla}}\ and\ \bibinfo {author} {\bibfnamefont {G.}~\bibnamefont
  {Torlai}},\ }\bibfield  {title} {\bibinfo {title} {How to use neural networks
  to investigate quantum many-body physics},\ }\href
  {https://doi.org/10.1103/PRXQuantum.2.040201} {\bibfield  {journal} {\bibinfo
   {journal} {PRX Quantum}\ }\textbf {\bibinfo {volume} {2}},\ \bibinfo {pages}
  {040201} (\bibinfo {year} {2021})}\BibitemShut {NoStop}%
\bibitem [{\citenamefont {Torlai}\ \emph {et~al.}(2023)\citenamefont {Torlai},
  \citenamefont {Wood}, \citenamefont {Acharya}, \citenamefont {Carleo},
  \citenamefont {Carrasquilla},\ and\ \citenamefont {Aolita}}]{qpt}%
  \BibitemOpen
  \bibfield  {author} {\bibinfo {author} {\bibfnamefont {G.}~\bibnamefont
  {Torlai}}, \bibinfo {author} {\bibfnamefont {C.~J.}\ \bibnamefont {Wood}},
  \bibinfo {author} {\bibfnamefont {A.}~\bibnamefont {Acharya}}, \bibinfo
  {author} {\bibfnamefont {G.}~\bibnamefont {Carleo}}, \bibinfo {author}
  {\bibfnamefont {J.}~\bibnamefont {Carrasquilla}},\ and\ \bibinfo {author}
  {\bibfnamefont {L.}~\bibnamefont {Aolita}},\ }\bibfield  {title} {\bibinfo
  {title} {Quantum process tomography with unsupervised learning and tensor
  networks},\ }\href {https://doi.org/10.1038/s41467-023-38332-9} {\bibfield
  {journal} {\bibinfo  {journal} {Nature Communications}\ }\textbf {\bibinfo
  {volume} {14}},\ \bibinfo {pages} {2858} (\bibinfo {year}
  {2023})}\BibitemShut {NoStop}%
\bibitem [{\citenamefont {Banchi}\ \emph {et~al.}(2018)\citenamefont {Banchi},
  \citenamefont {Grant}, \citenamefont {Rocchetto},\ and\ \citenamefont
  {Severini}}]{nm_rnn}%
  \BibitemOpen
  \bibfield  {author} {\bibinfo {author} {\bibfnamefont {L.}~\bibnamefont
  {Banchi}}, \bibinfo {author} {\bibfnamefont {E.}~\bibnamefont {Grant}},
  \bibinfo {author} {\bibfnamefont {A.}~\bibnamefont {Rocchetto}},\ and\
  \bibinfo {author} {\bibfnamefont {S.}~\bibnamefont {Severini}},\ }\bibfield
  {title} {\bibinfo {title} {Modelling non-markovian quantum processes with
  recurrent neural networks},\ }\href
  {https://doi.org/10.1088/1367-2630/aaf749} {\bibfield  {journal} {\bibinfo
  {journal} {New Journal of Physics}\ }\textbf {\bibinfo {volume} {20}},\
  \bibinfo {pages} {123030} (\bibinfo {year} {2018})}\BibitemShut {NoStop}%
\bibitem [{\citenamefont {Brunton}\ \emph {et~al.}(2016)\citenamefont
  {Brunton}, \citenamefont {Proctor},\ and\ \citenamefont {Kutz}}]{sindy}%
  \BibitemOpen
  \bibfield  {author} {\bibinfo {author} {\bibfnamefont {S.~L.}\ \bibnamefont
  {Brunton}}, \bibinfo {author} {\bibfnamefont {J.~L.}\ \bibnamefont
  {Proctor}},\ and\ \bibinfo {author} {\bibfnamefont {J.~N.}\ \bibnamefont
  {Kutz}},\ }\bibfield  {title} {\bibinfo {title} {Discovering governing
  equations from data by sparse identification of nonlinear dynamical
  systems},\ }\href {https://doi.org/10.1073/pnas.1517384113} {\bibfield
  {journal} {\bibinfo  {journal} {Proceedings of the National Academy of
  Sciences}\ }\textbf {\bibinfo {volume} {113}},\ \bibinfo {pages} {3932}
  (\bibinfo {year} {2016})}\BibitemShut {NoStop}%
\bibitem [{\citenamefont {Champion}\ \emph {et~al.}(2019)\citenamefont
  {Champion}, \citenamefont {Lusch}, \citenamefont {Kutz},\ and\ \citenamefont
  {Brunton}}]{sindy_champion}%
  \BibitemOpen
  \bibfield  {author} {\bibinfo {author} {\bibfnamefont {K.}~\bibnamefont
  {Champion}}, \bibinfo {author} {\bibfnamefont {B.}~\bibnamefont {Lusch}},
  \bibinfo {author} {\bibfnamefont {J.~N.}\ \bibnamefont {Kutz}},\ and\
  \bibinfo {author} {\bibfnamefont {S.~L.}\ \bibnamefont {Brunton}},\
  }\bibfield  {title} {\bibinfo {title} {Data-driven discovery of coordinates
  and governing equations},\ }\href {https://doi.org/10.1073/pnas.1906995116}
  {\bibfield  {journal} {\bibinfo  {journal} {Proceedings of the National
  Academy of Sciences}\ }\textbf {\bibinfo {volume} {116}},\ \bibinfo {pages}
  {22445} (\bibinfo {year} {2019})}\BibitemShut {NoStop}%
\bibitem [{\citenamefont {Lusch}\ \emph {et~al.}(2018)\citenamefont {Lusch},
  \citenamefont {Kutz},\ and\ \citenamefont {Brunton}}]{piml_encode}%
  \BibitemOpen
  \bibfield  {author} {\bibinfo {author} {\bibfnamefont {B.}~\bibnamefont
  {Lusch}}, \bibinfo {author} {\bibfnamefont {J.~N.}\ \bibnamefont {Kutz}},\
  and\ \bibinfo {author} {\bibfnamefont {S.~L.}\ \bibnamefont {Brunton}},\
  }\bibfield  {title} {\bibinfo {title} {Deep learning for universal linear
  embeddings of nonlinear dynamics},\ }\href
  {https://doi.org/10.1038/s41467-018-07210-0} {\bibfield  {journal} {\bibinfo
  {journal} {Nature Communications}\ }\textbf {\bibinfo {volume} {9}},\
  \bibinfo {pages} {4950} (\bibinfo {year} {2018})}\BibitemShut {NoStop}%
\bibitem [{\citenamefont {Genois}\ \emph {et~al.}(2021)\citenamefont {Genois},
  \citenamefont {Gross}, \citenamefont {Di~Paolo}, \citenamefont {Stevenson},
  \citenamefont {Koolstra}, \citenamefont {Hashim}, \citenamefont {Siddiqi},\
  and\ \citenamefont {Blais}}]{PRXQuantum.2.040355}%
  \BibitemOpen
  \bibfield  {author} {\bibinfo {author} {\bibfnamefont {E.}~\bibnamefont
  {Genois}}, \bibinfo {author} {\bibfnamefont {J.~A.}\ \bibnamefont {Gross}},
  \bibinfo {author} {\bibfnamefont {A.}~\bibnamefont {Di~Paolo}}, \bibinfo
  {author} {\bibfnamefont {N.~J.}\ \bibnamefont {Stevenson}}, \bibinfo {author}
  {\bibfnamefont {G.}~\bibnamefont {Koolstra}}, \bibinfo {author}
  {\bibfnamefont {A.}~\bibnamefont {Hashim}}, \bibinfo {author} {\bibfnamefont
  {I.}~\bibnamefont {Siddiqi}},\ and\ \bibinfo {author} {\bibfnamefont
  {A.}~\bibnamefont {Blais}},\ }\bibfield  {title} {\bibinfo {title}
  {Quantum-tailored machine-learning characterization of a superconducting
  qubit},\ }\href {https://doi.org/10.1103/PRXQuantum.2.040355} {\bibfield
  {journal} {\bibinfo  {journal} {PRX Quantum}\ }\textbf {\bibinfo {volume}
  {2}},\ \bibinfo {pages} {040355} (\bibinfo {year} {2021})}\BibitemShut
  {NoStop}%
\bibitem [{\citenamefont {Flurin}\ \emph {et~al.}(2020)\citenamefont {Flurin},
  \citenamefont {Martin}, \citenamefont {Hacohen-Gourgy},\ and\ \citenamefont
  {Siddiqi}}]{siddiqi_rnn}%
  \BibitemOpen
  \bibfield  {author} {\bibinfo {author} {\bibfnamefont {E.}~\bibnamefont
  {Flurin}}, \bibinfo {author} {\bibfnamefont {L.~S.}\ \bibnamefont {Martin}},
  \bibinfo {author} {\bibfnamefont {S.}~\bibnamefont {Hacohen-Gourgy}},\ and\
  \bibinfo {author} {\bibfnamefont {I.}~\bibnamefont {Siddiqi}},\ }\bibfield
  {title} {\bibinfo {title} {Using a recurrent neural network to reconstruct
  quantum dynamics of a superconducting qubit from physical observations},\
  }\href {https://doi.org/10.1103/PhysRevX.10.011006} {\bibfield  {journal}
  {\bibinfo  {journal} {Phys. Rev. X}\ }\textbf {\bibinfo {volume} {10}},\
  \bibinfo {pages} {011006} (\bibinfo {year} {2020})}\BibitemShut {NoStop}%
\bibitem [{\citenamefont {Koolstra}\ \emph {et~al.}(2022)\citenamefont
  {Koolstra}, \citenamefont {Stevenson}, \citenamefont {Barzili}, \citenamefont
  {Burns}, \citenamefont {Siva}, \citenamefont {Greenfield}, \citenamefont
  {Livingston}, \citenamefont {Hashim}, \citenamefont {Naik}, \citenamefont
  {Kreikebaum}, \citenamefont {O'Brien}, \citenamefont {Santiago},
  \citenamefont {Dressel},\ and\ \citenamefont {Siddiqi}}]{siddiqi_rnn2}%
  \BibitemOpen
  \bibfield  {author} {\bibinfo {author} {\bibfnamefont {G.}~\bibnamefont
  {Koolstra}}, \bibinfo {author} {\bibfnamefont {N.}~\bibnamefont {Stevenson}},
  \bibinfo {author} {\bibfnamefont {S.}~\bibnamefont {Barzili}}, \bibinfo
  {author} {\bibfnamefont {L.}~\bibnamefont {Burns}}, \bibinfo {author}
  {\bibfnamefont {K.}~\bibnamefont {Siva}}, \bibinfo {author} {\bibfnamefont
  {S.}~\bibnamefont {Greenfield}}, \bibinfo {author} {\bibfnamefont
  {W.}~\bibnamefont {Livingston}}, \bibinfo {author} {\bibfnamefont
  {A.}~\bibnamefont {Hashim}}, \bibinfo {author} {\bibfnamefont {R.~K.}\
  \bibnamefont {Naik}}, \bibinfo {author} {\bibfnamefont {J.~M.}\ \bibnamefont
  {Kreikebaum}}, \bibinfo {author} {\bibfnamefont {K.~P.}\ \bibnamefont
  {O'Brien}}, \bibinfo {author} {\bibfnamefont {D.~I.}\ \bibnamefont
  {Santiago}}, \bibinfo {author} {\bibfnamefont {J.}~\bibnamefont {Dressel}},\
  and\ \bibinfo {author} {\bibfnamefont {I.}~\bibnamefont {Siddiqi}},\
  }\bibfield  {title} {\bibinfo {title} {Monitoring fast superconducting qubit
  dynamics using a neural network},\ }\href
  {https://doi.org/10.1103/PhysRevX.12.031017} {\bibfield  {journal} {\bibinfo
  {journal} {Phys. Rev. X}\ }\textbf {\bibinfo {volume} {12}},\ \bibinfo
  {pages} {031017} (\bibinfo {year} {2022})}\BibitemShut {NoStop}%
\bibitem [{\citenamefont {An}\ \emph {et~al.}(2024)\citenamefont {An},
  \citenamefont {Wu}, \citenamefont {Yang}, \citenamefont {Zhou},\ and\
  \citenamefont {Zeng}}]{zan_trans}%
  \BibitemOpen
  \bibfield  {author} {\bibinfo {author} {\bibfnamefont {Z.}~\bibnamefont
  {An}}, \bibinfo {author} {\bibfnamefont {J.}~\bibnamefont {Wu}}, \bibinfo
  {author} {\bibfnamefont {M.}~\bibnamefont {Yang}}, \bibinfo {author}
  {\bibfnamefont {D.~L.}\ \bibnamefont {Zhou}},\ and\ \bibinfo {author}
  {\bibfnamefont {B.}~\bibnamefont {Zeng}},\ }\bibfield  {title} {\bibinfo
  {title} {Unified quantum state tomography and hamiltonian learning: A
  language-translation-like approach for quantum systems},\ }\href
  {https://doi.org/10.1103/PhysRevApplied.21.014037} {\bibfield  {journal}
  {\bibinfo  {journal} {Phys. Rev. Appl.}\ }\textbf {\bibinfo {volume} {21}},\
  \bibinfo {pages} {014037} (\bibinfo {year} {2024})}\BibitemShut {NoStop}%
\bibitem [{\citenamefont {Castelano}\ \emph {et~al.}(2024)\citenamefont
  {Castelano}, \citenamefont {Cunha}, \citenamefont {Luiz}, \citenamefont
  {de~Jesus~Napolitano}, \citenamefont {Prado},\ and\ \citenamefont
  {Fanchini}}]{castelano_piml}%
  \BibitemOpen
  \bibfield  {author} {\bibinfo {author} {\bibfnamefont {L.~K.}\ \bibnamefont
  {Castelano}}, \bibinfo {author} {\bibfnamefont {I.}~\bibnamefont {Cunha}},
  \bibinfo {author} {\bibfnamefont {F.~S.}\ \bibnamefont {Luiz}}, \bibinfo
  {author} {\bibfnamefont {R.}~\bibnamefont {de~Jesus~Napolitano}}, \bibinfo
  {author} {\bibfnamefont {M.~V. d.~S.}\ \bibnamefont {Prado}},\ and\ \bibinfo
  {author} {\bibfnamefont {F.~F.}\ \bibnamefont {Fanchini}},\ }\bibfield
  {title} {\bibinfo {title} {Combining physics-informed neural networks with
  the freezing mechanism for general hamiltonian learning},\ }\href
  {https://doi.org/10.1103/PhysRevA.110.032607} {\bibfield  {journal} {\bibinfo
   {journal} {Phys. Rev. A}\ }\textbf {\bibinfo {volume} {110}},\ \bibinfo
  {pages} {032607} (\bibinfo {year} {2024})}\BibitemShut {NoStop}%
\bibitem [{\citenamefont {Raissi}\ \emph {et~al.}(2019)\citenamefont {Raissi},
  \citenamefont {Perdikaris},\ and\ \citenamefont
  {Karniadakis}}]{Raissi_PIML2}%
  \BibitemOpen
  \bibfield  {author} {\bibinfo {author} {\bibfnamefont {M.}~\bibnamefont
  {Raissi}}, \bibinfo {author} {\bibfnamefont {P.}~\bibnamefont {Perdikaris}},\
  and\ \bibinfo {author} {\bibfnamefont {G.}~\bibnamefont {Karniadakis}},\
  }\bibfield  {title} {\bibinfo {title} {Physics-informed neural networks: A
  deep learning framework for solving forward and inverse problems involving
  nonlinear partial differential equations},\ }\href
  {https://doi.org/https://doi.org/10.1016/j.jcp.2018.10.045} {\bibfield
  {journal} {\bibinfo  {journal} {Journal of Computational Physics}\ }\textbf
  {\bibinfo {volume} {378}},\ \bibinfo {pages} {686} (\bibinfo {year}
  {2019})}\BibitemShut {NoStop}%
\bibitem [{\citenamefont {Chen}\ \emph {et~al.}(2018)\citenamefont {Chen},
  \citenamefont {Rubanova}, \citenamefont {Bettencourt},\ and\ \citenamefont
  {Duvenaud}}]{neural_ode}%
  \BibitemOpen
  \bibfield  {author} {\bibinfo {author} {\bibfnamefont {R.~T.~Q.}\
  \bibnamefont {Chen}}, \bibinfo {author} {\bibfnamefont {Y.}~\bibnamefont
  {Rubanova}}, \bibinfo {author} {\bibfnamefont {J.}~\bibnamefont
  {Bettencourt}},\ and\ \bibinfo {author} {\bibfnamefont {D.~K.}\ \bibnamefont
  {Duvenaud}},\ }\bibfield  {title} {\bibinfo {title} {Neural ordinary
  differential equations},\ }in\ \href
  {https://proceedings.neurips.cc/paper_files/paper/2018/file/69386f6bb1dfed68692a24c8686939b9-Paper.pdf}
  {\emph {\bibinfo {booktitle} {Advances in Neural Information Processing
  Systems}}},\ Vol.~\bibinfo {volume} {31},\ \bibinfo {editor} {edited by\
  \bibinfo {editor} {\bibfnamefont {S.}~\bibnamefont {Bengio}}, \bibinfo
  {editor} {\bibfnamefont {H.}~\bibnamefont {Wallach}}, \bibinfo {editor}
  {\bibfnamefont {H.}~\bibnamefont {Larochelle}}, \bibinfo {editor}
  {\bibfnamefont {K.}~\bibnamefont {Grauman}}, \bibinfo {editor} {\bibfnamefont
  {N.}~\bibnamefont {Cesa-Bianchi}},\ and\ \bibinfo {editor} {\bibfnamefont
  {R.}~\bibnamefont {Garnett}}}\ (\bibinfo  {publisher} {Curran Associates,
  Inc.},\ \bibinfo {year} {2018})\BibitemShut {NoStop}%
\bibitem [{\citenamefont {Habiba}\ and\ \citenamefont
  {Pearlmutter}(2020)}]{odernn}%
  \BibitemOpen
  \bibfield  {author} {\bibinfo {author} {\bibfnamefont {M.}~\bibnamefont
  {Habiba}}\ and\ \bibinfo {author} {\bibfnamefont {B.~A.}\ \bibnamefont
  {Pearlmutter}},\ }\bibfield  {title} {\bibinfo {title} {Neural ordinary
  differential equation based recurrent neural network model},\ }in\ \href
  {https://doi.org/10.1109/ISSC49989.2020.9180182} {\emph {\bibinfo {booktitle}
  {2020 31st Irish Signals and Systems Conference (ISSC)}}}\ (\bibinfo {year}
  {2020})\ pp.\ \bibinfo {pages} {1--6}\BibitemShut {NoStop}%
\bibitem [{\citenamefont {Smolin}\ \emph {et~al.}(2012)\citenamefont {Smolin},
  \citenamefont {Gambetta},\ and\ \citenamefont {Smith}}]{proj_density}%
  \BibitemOpen
  \bibfield  {author} {\bibinfo {author} {\bibfnamefont {J.~A.}\ \bibnamefont
  {Smolin}}, \bibinfo {author} {\bibfnamefont {J.~M.}\ \bibnamefont
  {Gambetta}},\ and\ \bibinfo {author} {\bibfnamefont {G.}~\bibnamefont
  {Smith}},\ }\bibfield  {title} {\bibinfo {title} {Efficient method for
  computing the maximum-likelihood quantum state from measurements with
  additive gaussian noise},\ }\href
  {https://doi.org/10.1103/PhysRevLett.108.070502} {\bibfield  {journal}
  {\bibinfo  {journal} {Phys. Rev. Lett.}\ }\textbf {\bibinfo {volume} {108}},\
  \bibinfo {pages} {070502} (\bibinfo {year} {2012})}\BibitemShut {NoStop}%
\bibitem [{\citenamefont {Kaheman}\ \emph {et~al.}(2019)\citenamefont
  {Kaheman}, \citenamefont {Kaiser}, \citenamefont {Strom}, \citenamefont
  {Kutz},\ and\ \citenamefont {Brunton}}]{disc}%
  \BibitemOpen
  \bibfield  {author} {\bibinfo {author} {\bibfnamefont {K.}~\bibnamefont
  {Kaheman}}, \bibinfo {author} {\bibfnamefont {E.}~\bibnamefont {Kaiser}},
  \bibinfo {author} {\bibfnamefont {B.}~\bibnamefont {Strom}}, \bibinfo
  {author} {\bibfnamefont {J.~N.}\ \bibnamefont {Kutz}},\ and\ \bibinfo
  {author} {\bibfnamefont {S.~L.}\ \bibnamefont {Brunton}},\ }\href@noop {}
  {\bibinfo {title} {Learning discrepancy models from experimental data}}
  (\bibinfo {year} {2019}),\ \Eprint {https://arxiv.org/abs/1909.08574}
  {arXiv:1909.08574 [cs.LG]} \BibitemShut {NoStop}%
\bibitem [{\citenamefont {de~Silva}\ \emph {et~al.}(2020)\citenamefont
  {de~Silva}, \citenamefont {Higdon}, \citenamefont {Brunton},\ and\
  \citenamefont {Kutz}}]{disc2}%
  \BibitemOpen
  \bibfield  {author} {\bibinfo {author} {\bibfnamefont {B.~M.}\ \bibnamefont
  {de~Silva}}, \bibinfo {author} {\bibfnamefont {D.~M.}\ \bibnamefont
  {Higdon}}, \bibinfo {author} {\bibfnamefont {S.~L.}\ \bibnamefont
  {Brunton}},\ and\ \bibinfo {author} {\bibfnamefont {J.~N.}\ \bibnamefont
  {Kutz}},\ }\bibfield  {title} {\bibinfo {title} {Discovery of physics from
  data: Universal laws and discrepancies},\ }\bibfield  {journal} {\bibinfo
  {journal} {Frontiers in Artificial Intelligence}\ }\textbf {\bibinfo {volume}
  {3}},\ \href {https://doi.org/10.3389/frai.2020.00025}
  {10.3389/frai.2020.00025} (\bibinfo {year} {2020})\BibitemShut {NoStop}%
\bibitem [{\citenamefont {Goodfellow}\ \emph {et~al.}(2016)\citenamefont
  {Goodfellow}, \citenamefont {Bengio},\ and\ \citenamefont
  {Courville}}]{goodfellow}%
  \BibitemOpen
  \bibfield  {author} {\bibinfo {author} {\bibfnamefont {I.}~\bibnamefont
  {Goodfellow}}, \bibinfo {author} {\bibfnamefont {Y.}~\bibnamefont {Bengio}},\
  and\ \bibinfo {author} {\bibfnamefont {A.}~\bibnamefont {Courville}},\
  }\href@noop {} {\emph {\bibinfo {title} {Deep Learning}}}\ (\bibinfo
  {publisher} {MIT Press},\ \bibinfo {year} {2016})\ \bibinfo {note}
  {\url{http://www.deeplearningbook.org}}\BibitemShut {NoStop}%
\bibitem [{\citenamefont {Haah}\ \emph {et~al.}(2024)\citenamefont {Haah},
  \citenamefont {Kothari},\ and\ \citenamefont {Tang}}]{Haah2024}%
  \BibitemOpen
  \bibfield  {author} {\bibinfo {author} {\bibfnamefont {J.}~\bibnamefont
  {Haah}}, \bibinfo {author} {\bibfnamefont {R.}~\bibnamefont {Kothari}},\ and\
  \bibinfo {author} {\bibfnamefont {E.}~\bibnamefont {Tang}},\ }\bibfield
  {title} {\bibinfo {title} {Learning quantum hamiltonians from
  high-temperature gibbs states and real-time evolutions},\ }\href
  {https://doi.org/10.1038/s41567-023-02376-x} {\bibfield  {journal} {\bibinfo
  {journal} {Nature Physics}\ }\textbf {\bibinfo {volume} {20}},\ \bibinfo
  {pages} {1027} (\bibinfo {year} {2024})}\BibitemShut {NoStop}%
\bibitem [{\citenamefont {Anshu}\ \emph {et~al.}(2021)\citenamefont {Anshu},
  \citenamefont {Arunachalam}, \citenamefont {Kuwahara},\ and\ \citenamefont
  {Soleimanifar}}]{Anshu2021}%
  \BibitemOpen
  \bibfield  {author} {\bibinfo {author} {\bibfnamefont {A.}~\bibnamefont
  {Anshu}}, \bibinfo {author} {\bibfnamefont {S.}~\bibnamefont {Arunachalam}},
  \bibinfo {author} {\bibfnamefont {T.}~\bibnamefont {Kuwahara}},\ and\
  \bibinfo {author} {\bibfnamefont {M.}~\bibnamefont {Soleimanifar}},\
  }\bibfield  {title} {\bibinfo {title} {Sample-efficient learning of
  interacting quantum systems},\ }\href
  {https://doi.org/10.1038/s41567-021-01232-0} {\bibfield  {journal} {\bibinfo
  {journal} {Nature Physics}\ }\textbf {\bibinfo {volume} {17}},\ \bibinfo
  {pages} {931} (\bibinfo {year} {2021})}\BibitemShut {NoStop}%
\bibitem [{\citenamefont {Anshu}\ and\ \citenamefont
  {Arunachalam}(2024)}]{Anshu2024}%
  \BibitemOpen
  \bibfield  {author} {\bibinfo {author} {\bibfnamefont {A.}~\bibnamefont
  {Anshu}}\ and\ \bibinfo {author} {\bibfnamefont {S.}~\bibnamefont
  {Arunachalam}},\ }\bibfield  {title} {\bibinfo {title} {A survey on the
  complexity of learning quantum states},\ }\href
  {https://doi.org/10.1038/s42254-023-00662-4} {\bibfield  {journal} {\bibinfo
  {journal} {Nature Reviews Physics}\ }\textbf {\bibinfo {volume} {6}},\
  \bibinfo {pages} {59} (\bibinfo {year} {2024})}\BibitemShut {NoStop}%
\bibitem [{\citenamefont {Huang}\ \emph {et~al.}(2023)\citenamefont {Huang},
  \citenamefont {Tong}, \citenamefont {Fang},\ and\ \citenamefont
  {Su}}]{PhysRevLett.130.200403}%
  \BibitemOpen
  \bibfield  {author} {\bibinfo {author} {\bibfnamefont {H.-Y.}\ \bibnamefont
  {Huang}}, \bibinfo {author} {\bibfnamefont {Y.}~\bibnamefont {Tong}},
  \bibinfo {author} {\bibfnamefont {D.}~\bibnamefont {Fang}},\ and\ \bibinfo
  {author} {\bibfnamefont {Y.}~\bibnamefont {Su}},\ }\bibfield  {title}
  {\bibinfo {title} {Learning many-body hamiltonians with heisenberg-limited
  scaling},\ }\href {https://doi.org/10.1103/PhysRevLett.130.200403} {\bibfield
   {journal} {\bibinfo  {journal} {Phys. Rev. Lett.}\ }\textbf {\bibinfo
  {volume} {130}},\ \bibinfo {pages} {200403} (\bibinfo {year}
  {2023})}\BibitemShut {NoStop}%
\bibitem [{\citenamefont {Li}\ \emph {et~al.}(2023)\citenamefont {Li},
  \citenamefont {Tong}, \citenamefont {Ni}, \citenamefont {Gefen},\ and\
  \citenamefont {Ying}}]{arxiv2307.04690}%
  \BibitemOpen
  \bibfield  {author} {\bibinfo {author} {\bibfnamefont {H.}~\bibnamefont
  {Li}}, \bibinfo {author} {\bibfnamefont {Y.}~\bibnamefont {Tong}}, \bibinfo
  {author} {\bibfnamefont {H.}~\bibnamefont {Ni}}, \bibinfo {author}
  {\bibfnamefont {T.}~\bibnamefont {Gefen}},\ and\ \bibinfo {author}
  {\bibfnamefont {L.}~\bibnamefont {Ying}},\ }\href
  {https://arxiv.org/abs/2307.04690} {\bibinfo {title} {Heisenberg-limited
  hamiltonian learning for interacting bosons}} (\bibinfo {year} {2023}),\
  \Eprint {https://arxiv.org/abs/2307.04690} {arXiv:2307.04690 [quant-ph]}
  \BibitemShut {NoStop}%
\bibitem [{\citenamefont {Dutkiewicz}\ \emph {et~al.}(2024)\citenamefont
  {Dutkiewicz}, \citenamefont {O'Brien},\ and\ \citenamefont
  {Schuster}}]{arxiv2304.07172}%
  \BibitemOpen
  \bibfield  {author} {\bibinfo {author} {\bibfnamefont {A.}~\bibnamefont
  {Dutkiewicz}}, \bibinfo {author} {\bibfnamefont {T.~E.}\ \bibnamefont
  {O'Brien}},\ and\ \bibinfo {author} {\bibfnamefont {T.}~\bibnamefont
  {Schuster}},\ }\href {https://arxiv.org/abs/2304.07172} {\bibinfo {title}
  {The advantage of quantum control in many-body hamiltonian learning}}
  (\bibinfo {year} {2024}),\ \Eprint {https://arxiv.org/abs/2304.07172}
  {arXiv:2304.07172 [quant-ph]} \BibitemShut {NoStop}%
\bibitem [{\citenamefont {Bairey}\ \emph {et~al.}(2019)\citenamefont {Bairey},
  \citenamefont {Arad},\ and\ \citenamefont
  {Lindner}}]{PhysRevLett.122.020504}%
  \BibitemOpen
  \bibfield  {author} {\bibinfo {author} {\bibfnamefont {E.}~\bibnamefont
  {Bairey}}, \bibinfo {author} {\bibfnamefont {I.}~\bibnamefont {Arad}},\ and\
  \bibinfo {author} {\bibfnamefont {N.~H.}\ \bibnamefont {Lindner}},\
  }\bibfield  {title} {\bibinfo {title} {Learning a local hamiltonian from
  local measurements},\ }\href {https://doi.org/10.1103/PhysRevLett.122.020504}
  {\bibfield  {journal} {\bibinfo  {journal} {Phys. Rev. Lett.}\ }\textbf
  {\bibinfo {volume} {122}},\ \bibinfo {pages} {020504} (\bibinfo {year}
  {2019})}\BibitemShut {NoStop}%
\bibitem [{\citenamefont {Bairey}\ \emph {et~al.}(2020)\citenamefont {Bairey},
  \citenamefont {Guo}, \citenamefont {Poletti}, \citenamefont {Lindner},\ and\
  \citenamefont {Arad}}]{Bairey_2020}%
  \BibitemOpen
  \bibfield  {author} {\bibinfo {author} {\bibfnamefont {E.}~\bibnamefont
  {Bairey}}, \bibinfo {author} {\bibfnamefont {C.}~\bibnamefont {Guo}},
  \bibinfo {author} {\bibfnamefont {D.}~\bibnamefont {Poletti}}, \bibinfo
  {author} {\bibfnamefont {N.~H.}\ \bibnamefont {Lindner}},\ and\ \bibinfo
  {author} {\bibfnamefont {I.}~\bibnamefont {Arad}},\ }\bibfield  {title}
  {\bibinfo {title} {Learning the dynamics of open quantum systems from their
  steady states},\ }\href {https://doi.org/10.1088/1367-2630/ab73cd} {\bibfield
   {journal} {\bibinfo  {journal} {New Journal of Physics}\ }\textbf {\bibinfo
  {volume} {22}},\ \bibinfo {pages} {032001} (\bibinfo {year}
  {2020})}\BibitemShut {NoStop}%
\bibitem [{\citenamefont {Jacobs}\ and\ \citenamefont
  {Steck}(2006)}]{cont_meas}%
  \BibitemOpen
  \bibfield  {author} {\bibinfo {author} {\bibfnamefont {K.}~\bibnamefont
  {Jacobs}}\ and\ \bibinfo {author} {\bibfnamefont {D.~A.}\ \bibnamefont
  {Steck}},\ }\bibfield  {title} {\bibinfo {title} {A straightforward
  introduction to continuous quantum measurement},\ }\href
  {https://doi.org/10.1080/00107510601101934} {\bibfield  {journal} {\bibinfo
  {journal} {Contemporary Physics}\ }\textbf {\bibinfo {volume} {47}},\
  \bibinfo {pages} {279} (\bibinfo {year} {2006})}\BibitemShut {NoStop}%
\bibitem [{\citenamefont {Brun}(2002)}]{brun}%
  \BibitemOpen
  \bibfield  {author} {\bibinfo {author} {\bibfnamefont {T.~A.}\ \bibnamefont
  {Brun}},\ }\bibfield  {title} {\bibinfo {title} {{A simple model of quantum
  trajectories}},\ }\href {https://doi.org/10.1119/1.1475328} {\bibfield
  {journal} {\bibinfo  {journal} {American Journal of Physics}\ }\textbf
  {\bibinfo {volume} {70}},\ \bibinfo {pages} {719} (\bibinfo {year}
  {2002})}\BibitemShut {NoStop}%
\bibitem [{\citenamefont {Bonifacio}\ \emph {et~al.}(1971)\citenamefont
  {Bonifacio}, \citenamefont {Schwendimann},\ and\ \citenamefont
  {Haake}}]{Haake_1971}%
  \BibitemOpen
  \bibfield  {author} {\bibinfo {author} {\bibfnamefont {R.}~\bibnamefont
  {Bonifacio}}, \bibinfo {author} {\bibfnamefont {P.}~\bibnamefont
  {Schwendimann}},\ and\ \bibinfo {author} {\bibfnamefont {F.}~\bibnamefont
  {Haake}},\ }\bibfield  {title} {\bibinfo {title} {Quantum statistical theory
  of superradiance. i},\ }\href {https://doi.org/10.1103/PhysRevA.4.302}
  {\bibfield  {journal} {\bibinfo  {journal} {Phys. Rev. A}\ }\textbf {\bibinfo
  {volume} {4}},\ \bibinfo {pages} {302} (\bibinfo {year} {1971})}\BibitemShut
  {NoStop}%
\bibitem [{\citenamefont {Blais}\ \emph {et~al.}(2021)\citenamefont {Blais},
  \citenamefont {Grimsmo}, \citenamefont {Girvin},\ and\ \citenamefont
  {Wallraff}}]{circuit_qed}%
  \BibitemOpen
  \bibfield  {author} {\bibinfo {author} {\bibfnamefont {A.}~\bibnamefont
  {Blais}}, \bibinfo {author} {\bibfnamefont {A.~L.}\ \bibnamefont {Grimsmo}},
  \bibinfo {author} {\bibfnamefont {S.~M.}\ \bibnamefont {Girvin}},\ and\
  \bibinfo {author} {\bibfnamefont {A.}~\bibnamefont {Wallraff}},\ }\bibfield
  {title} {\bibinfo {title} {Circuit quantum electrodynamics},\ }\href
  {https://doi.org/10.1103/RevModPhys.93.025005} {\bibfield  {journal}
  {\bibinfo  {journal} {Rev. Mod. Phys.}\ }\textbf {\bibinfo {volume} {93}},\
  \bibinfo {pages} {025005} (\bibinfo {year} {2021})}\BibitemShut {NoStop}%
\bibitem [{\citenamefont {Gambetta}\ \emph {et~al.}(2008)\citenamefont
  {Gambetta}, \citenamefont {Blais}, \citenamefont {Boissonneault},
  \citenamefont {Houck}, \citenamefont {Schuster},\ and\ \citenamefont
  {Girvin}}]{circuit_qed_traj}%
  \BibitemOpen
  \bibfield  {author} {\bibinfo {author} {\bibfnamefont {J.}~\bibnamefont
  {Gambetta}}, \bibinfo {author} {\bibfnamefont {A.}~\bibnamefont {Blais}},
  \bibinfo {author} {\bibfnamefont {M.}~\bibnamefont {Boissonneault}}, \bibinfo
  {author} {\bibfnamefont {A.~A.}\ \bibnamefont {Houck}}, \bibinfo {author}
  {\bibfnamefont {D.~I.}\ \bibnamefont {Schuster}},\ and\ \bibinfo {author}
  {\bibfnamefont {S.~M.}\ \bibnamefont {Girvin}},\ }\bibfield  {title}
  {\bibinfo {title} {Quantum trajectory approach to circuit qed: Quantum jumps
  and the zeno effect},\ }\href {https://doi.org/10.1103/PhysRevA.77.012112}
  {\bibfield  {journal} {\bibinfo  {journal} {Phys. Rev. A}\ }\textbf {\bibinfo
  {volume} {77}},\ \bibinfo {pages} {012112} (\bibinfo {year}
  {2008})}\BibitemShut {NoStop}%
\bibitem [{\citenamefont {Rumelhart}\ \emph {et~al.}(1986)\citenamefont
  {Rumelhart}, \citenamefont {Hinton},\ and\ \citenamefont
  {Williams}}]{rumelhart:errorpropnonote}%
  \BibitemOpen
  \bibfield  {author} {\bibinfo {author} {\bibfnamefont {D.~E.}\ \bibnamefont
  {Rumelhart}}, \bibinfo {author} {\bibfnamefont {G.~E.}\ \bibnamefont
  {Hinton}},\ and\ \bibinfo {author} {\bibfnamefont {R.~J.}\ \bibnamefont
  {Williams}},\ }\bibfield  {title} {\bibinfo {title} {Learning internal
  representations by error propagation},\ }in\ \href@noop {} {\emph {\bibinfo
  {booktitle} {Parallel Distributed Processing: Explorations in the
  Microstructure of Cognition, {V}olume 1: {F}oundations}}},\ \bibinfo {editor}
  {edited by\ \bibinfo {editor} {\bibfnamefont {D.~E.}\ \bibnamefont
  {Rumelhart}}\ and\ \bibinfo {editor} {\bibfnamefont {J.~L.}\ \bibnamefont
  {Mcclelland}}}\ (\bibinfo  {publisher} {MIT Press},\ \bibinfo {address}
  {Cambridge, MA},\ \bibinfo {year} {1986})\ pp.\ \bibinfo {pages}
  {318--362}\BibitemShut {NoStop}%
\bibitem [{\citenamefont {Hochreiter}\ \emph {et~al.}(2001)\citenamefont
  {Hochreiter}, \citenamefont {Bengio}, \citenamefont {Frasconi},\ and\
  \citenamefont {Schmidhuber}}]{Hochreiter:01book}%
  \BibitemOpen
  \bibfield  {author} {\bibinfo {author} {\bibfnamefont {S.}~\bibnamefont
  {Hochreiter}}, \bibinfo {author} {\bibfnamefont {Y.}~\bibnamefont {Bengio}},
  \bibinfo {author} {\bibfnamefont {P.}~\bibnamefont {Frasconi}},\ and\
  \bibinfo {author} {\bibfnamefont {J.}~\bibnamefont {Schmidhuber}},\
  }\bibfield  {title} {\bibinfo {title} {Gradient flow in recurrent nets: the
  difficulty of learning long-term dependencies},\ }in\ \href@noop {} {\emph
  {\bibinfo {booktitle} {A Field Guide to Dynamical Recurrent Neural
  Networks}}},\ \bibinfo {editor} {edited by\ \bibinfo {editor} {\bibfnamefont
  {S.~C.}\ \bibnamefont {Kremer}}\ and\ \bibinfo {editor} {\bibfnamefont
  {J.~F.}\ \bibnamefont {Kolen}}}\ (\bibinfo  {publisher} {IEEE Press},\
  \bibinfo {year} {2001})\BibitemShut {NoStop}%
\bibitem [{\citenamefont {Hochreiter}\ and\ \citenamefont
  {Schmidhuber}(1997)}]{lstm}%
  \BibitemOpen
  \bibfield  {author} {\bibinfo {author} {\bibfnamefont {S.}~\bibnamefont
  {Hochreiter}}\ and\ \bibinfo {author} {\bibfnamefont {J.}~\bibnamefont
  {Schmidhuber}},\ }\bibfield  {title} {\bibinfo {title} {Long short-term
  memory},\ }\href {https://doi.org/10.1162/neco.1997.9.8.1735} {\bibfield
  {journal} {\bibinfo  {journal} {Neural Comput.}\ }\textbf {\bibinfo {volume}
  {9}},\ \bibinfo {pages} {1735–1780} (\bibinfo {year} {1997})}\BibitemShut
  {NoStop}%
\bibitem [{\citenamefont {Rubanova}\ \emph {et~al.}(2019)\citenamefont
  {Rubanova}, \citenamefont {Chen},\ and\ \citenamefont
  {Duvenaud}}]{odernn_latent}%
  \BibitemOpen
  \bibfield  {author} {\bibinfo {author} {\bibfnamefont {Y.}~\bibnamefont
  {Rubanova}}, \bibinfo {author} {\bibfnamefont {R.~T.~Q.}\ \bibnamefont
  {Chen}},\ and\ \bibinfo {author} {\bibfnamefont {D.~K.}\ \bibnamefont
  {Duvenaud}},\ }\bibfield  {title} {\bibinfo {title} {Latent ordinary
  differential equations for irregularly-sampled time series},\ }in\ \href
  {https://proceedings.neurips.cc/paper_files/paper/2019/file/42a6845a557bef704ad8ac9cb4461d43-Paper.pdf}
  {\emph {\bibinfo {booktitle} {Advances in Neural Information Processing
  Systems}}},\ Vol.~\bibinfo {volume} {32},\ \bibinfo {editor} {edited by\
  \bibinfo {editor} {\bibfnamefont {H.}~\bibnamefont {Wallach}}, \bibinfo
  {editor} {\bibfnamefont {H.}~\bibnamefont {Larochelle}}, \bibinfo {editor}
  {\bibfnamefont {A.}~\bibnamefont {Beygelzimer}}, \bibinfo {editor}
  {\bibfnamefont {F.}~\bibnamefont {d\textquotesingle Alch\'{e}-Buc}}, \bibinfo
  {editor} {\bibfnamefont {E.}~\bibnamefont {Fox}},\ and\ \bibinfo {editor}
  {\bibfnamefont {R.}~\bibnamefont {Garnett}}}\ (\bibinfo  {publisher} {Curran
  Associates, Inc.},\ \bibinfo {year} {2019})\BibitemShut {NoStop}%
\bibitem [{\citenamefont {Haegeman}\ \emph {et~al.}(2011)\citenamefont
  {Haegeman}, \citenamefont {Cirac}, \citenamefont {Osborne}, \citenamefont
  {Pi\ifmmode~\check{z}\else \v{z}\fi{}orn}, \citenamefont {Verschelde},\ and\
  \citenamefont {Verstraete}}]{tdvp}%
  \BibitemOpen
  \bibfield  {author} {\bibinfo {author} {\bibfnamefont {J.}~\bibnamefont
  {Haegeman}}, \bibinfo {author} {\bibfnamefont {J.~I.}\ \bibnamefont {Cirac}},
  \bibinfo {author} {\bibfnamefont {T.~J.}\ \bibnamefont {Osborne}}, \bibinfo
  {author} {\bibfnamefont {I.}~\bibnamefont {Pi\ifmmode~\check{z}\else
  \v{z}\fi{}orn}}, \bibinfo {author} {\bibfnamefont {H.}~\bibnamefont
  {Verschelde}},\ and\ \bibinfo {author} {\bibfnamefont {F.}~\bibnamefont
  {Verstraete}},\ }\bibfield  {title} {\bibinfo {title} {Time-dependent
  variational principle for quantum lattices},\ }\href
  {https://doi.org/10.1103/PhysRevLett.107.070601} {\bibfield  {journal}
  {\bibinfo  {journal} {Phys. Rev. Lett.}\ }\textbf {\bibinfo {volume} {107}},\
  \bibinfo {pages} {070601} (\bibinfo {year} {2011})}\BibitemShut {NoStop}%
\bibitem [{\citenamefont {Haegeman}\ \emph {et~al.}(2016)\citenamefont
  {Haegeman}, \citenamefont {Lubich}, \citenamefont {Oseledets}, \citenamefont
  {Vandereycken},\ and\ \citenamefont {Verstraete}}]{tdvp2}%
  \BibitemOpen
  \bibfield  {author} {\bibinfo {author} {\bibfnamefont {J.}~\bibnamefont
  {Haegeman}}, \bibinfo {author} {\bibfnamefont {C.}~\bibnamefont {Lubich}},
  \bibinfo {author} {\bibfnamefont {I.}~\bibnamefont {Oseledets}}, \bibinfo
  {author} {\bibfnamefont {B.}~\bibnamefont {Vandereycken}},\ and\ \bibinfo
  {author} {\bibfnamefont {F.}~\bibnamefont {Verstraete}},\ }\bibfield  {title}
  {\bibinfo {title} {Unifying time evolution and optimization with matrix
  product states},\ }\href {https://doi.org/10.1103/PhysRevB.94.165116}
  {\bibfield  {journal} {\bibinfo  {journal} {Phys. Rev. B}\ }\textbf {\bibinfo
  {volume} {94}},\ \bibinfo {pages} {165116} (\bibinfo {year}
  {2016})}\BibitemShut {NoStop}%
\bibitem [{\citenamefont {Perez-Garcia}\ \emph {et~al.}(2007)\citenamefont
  {Perez-Garcia}, \citenamefont {Verstraete}, \citenamefont {Wolf},\ and\
  \citenamefont {Cirac}}]{perez2006matrix}%
  \BibitemOpen
  \bibfield  {author} {\bibinfo {author} {\bibfnamefont {D.}~\bibnamefont
  {Perez-Garcia}}, \bibinfo {author} {\bibfnamefont {F.}~\bibnamefont
  {Verstraete}}, \bibinfo {author} {\bibfnamefont {M.~M.}\ \bibnamefont
  {Wolf}},\ and\ \bibinfo {author} {\bibfnamefont {J.~I.}\ \bibnamefont
  {Cirac}},\ }\bibfield  {title} {\bibinfo {title} {Matrix product state
  representations},\ }\href {https://doi.org/10.48550/arXiv.quant-ph/0608197}
  {\bibfield  {journal} {\bibinfo  {journal} {Quantum Info. Comput.}\ }\textbf
  {\bibinfo {volume} {7}},\ \bibinfo {pages} {401} (\bibinfo {year}
  {2007})}\BibitemShut {NoStop}%
\bibitem [{\citenamefont {Schollw{\"o}ck}(2011)}]{schollwock2011density}%
  \BibitemOpen
  \bibfield  {author} {\bibinfo {author} {\bibfnamefont {U.}~\bibnamefont
  {Schollw{\"o}ck}},\ }\bibfield  {title} {\bibinfo {title} {The density-matrix
  renormalization group in the age of matrix product states},\ }\href
  {https://doi.org/https://doi.org/10.1016/j.aop.2010.09.012} {\bibfield
  {journal} {\bibinfo  {journal} {Annals of Physics}\ }\textbf {\bibinfo
  {volume} {326}},\ \bibinfo {pages} {96} (\bibinfo {year} {2011})}\BibitemShut
  {NoStop}%
\bibitem [{\citenamefont {Bultink}\ \emph {et~al.}(2018)\citenamefont
  {Bultink}, \citenamefont {Tarasinski}, \citenamefont {Haandbæk},
  \citenamefont {Poletto}, \citenamefont {Haider}, \citenamefont {Michalak},
  \citenamefont {Bruno},\ and\ \citenamefont {DiCarlo}}]{eta_est}%
  \BibitemOpen
  \bibfield  {author} {\bibinfo {author} {\bibfnamefont {C.~C.}\ \bibnamefont
  {Bultink}}, \bibinfo {author} {\bibfnamefont {B.}~\bibnamefont {Tarasinski}},
  \bibinfo {author} {\bibfnamefont {N.}~\bibnamefont {Haandbæk}}, \bibinfo
  {author} {\bibfnamefont {S.}~\bibnamefont {Poletto}}, \bibinfo {author}
  {\bibfnamefont {N.}~\bibnamefont {Haider}}, \bibinfo {author} {\bibfnamefont
  {D.~J.}\ \bibnamefont {Michalak}}, \bibinfo {author} {\bibfnamefont
  {A.}~\bibnamefont {Bruno}},\ and\ \bibinfo {author} {\bibfnamefont
  {L.}~\bibnamefont {DiCarlo}},\ }\bibfield  {title} {\bibinfo {title}
  {{General method for extracting the quantum efficiency of dispersive qubit
  readout in circuit QED}},\ }\href {https://doi.org/10.1063/1.5015954}
  {\bibfield  {journal} {\bibinfo  {journal} {Applied Physics Letters}\
  }\textbf {\bibinfo {volume} {112}},\ \bibinfo {pages} {092601} (\bibinfo
  {year} {2018})},\ \Eprint
  {https://arxiv.org/abs/https://pubs.aip.org/aip/apl/article-pdf/doi/10.1063/1.5015954/13033954/092601\_1\_online.pdf}
  {https://pubs.aip.org/aip/apl/article-pdf/doi/10.1063/1.5015954/13033954/092601\_1\_online.pdf}
  \BibitemShut {NoStop}%
\bibitem [{\citenamefont {Jurafsky}\ and\ \citenamefont
  {Martin}(2024)}]{jurafskyMartin24}%
  \BibitemOpen
  \bibfield  {author} {\bibinfo {author} {\bibfnamefont {D.}~\bibnamefont
  {Jurafsky}}\ and\ \bibinfo {author} {\bibfnamefont {J.~H.}\ \bibnamefont
  {Martin}},\ }\href@noop {} {\emph {\bibinfo {title} {Speech and Language
  Processing An Introduction to Natural Language Processing, Computational
  Linguistics, and Speech Recognition}}},\ \bibinfo {edition} {3rd}\ ed.\
  (\bibinfo {year} {2024})\BibitemShut {NoStop}%
\bibitem [{\citenamefont {Elman}(1990)}]{elman_rnn}%
  \BibitemOpen
  \bibfield  {author} {\bibinfo {author} {\bibfnamefont {J.~L.}\ \bibnamefont
  {Elman}},\ }\bibfield  {title} {\bibinfo {title} {Finding structure in
  time},\ }\href {https://doi.org/https://doi.org/10.1207/s15516709cog1402\_1}
  {\bibfield  {journal} {\bibinfo  {journal} {Cognitive Science}\ }\textbf
  {\bibinfo {volume} {14}},\ \bibinfo {pages} {179} (\bibinfo {year}
  {1990})}\BibitemShut {NoStop}%
\bibitem [{\citenamefont {Robinson}\ \emph {et~al.}(1987)\citenamefont
  {Robinson}, \citenamefont {Fallside},\ and\ \citenamefont {of~Cambridge.
  Engineering~Department}}]{Robinson1987}%
  \BibitemOpen
  \bibfield  {author} {\bibinfo {author} {\bibfnamefont {A.}~\bibnamefont
  {Robinson}}, \bibinfo {author} {\bibfnamefont {F.}~\bibnamefont {Fallside}},\
  and\ \bibinfo {author} {\bibfnamefont {U.}~\bibnamefont {of~Cambridge.
  Engineering~Department}},\ }\href
  {https://books.google.com/books?id=6JYYMwEACAAJ} {\emph {\bibinfo {title}
  {The Utility Driven Dynamic Error Propagation Network}}}\ (\bibinfo
  {publisher} {University of Cambridge Department of Engineering},\ \bibinfo
  {year} {1987})\BibitemShut {NoStop}%
\bibitem [{\citenamefont {Werbos}(1988)}]{Werbos1988}%
  \BibitemOpen
  \bibfield  {author} {\bibinfo {author} {\bibfnamefont {P.~J.}\ \bibnamefont
  {Werbos}},\ }\bibfield  {title} {\bibinfo {title} {Generalization of
  backpropagation with application to a recurrent gas market model},\ }\href
  {https://doi.org/https://doi.org/10.1016/0893-6080(88)90007-X} {\bibfield
  {journal} {\bibinfo  {journal} {Neural Networks}\ }\textbf {\bibinfo {volume}
  {1}},\ \bibinfo {pages} {339} (\bibinfo {year} {1988})}\BibitemShut {NoStop}%
\bibitem [{\citenamefont {Mozer}(1995)}]{Mozer1995}%
  \BibitemOpen
  \bibfield  {author} {\bibinfo {author} {\bibfnamefont {M.~C.}\ \bibnamefont
  {Mozer}},\ }\bibinfo {title} {A focused backpropagation algorithm for
  temporal pattern recognition},\ in\ \href@noop {} {\emph {\bibinfo
  {booktitle} {Backpropagation: Theory, Architectures, and Applications}}}\
  (\bibinfo  {publisher} {L. Erlbaum Associates Inc.},\ \bibinfo {address}
  {USA},\ \bibinfo {year} {1995})\ p.\ \bibinfo {pages} {137–169}\BibitemShut
  {NoStop}%
\bibitem [{\citenamefont {Breiman}(2001)}]{breiman2001random}%
  \BibitemOpen
  \bibfield  {author} {\bibinfo {author} {\bibfnamefont {L.}~\bibnamefont
  {Breiman}},\ }\bibfield  {title} {\bibinfo {title} {Random forests},\ }\href
  {https://doi.org/10.1023/A:1010933404324} {\bibfield  {journal} {\bibinfo
  {journal} {Machine Learning}\ }\textbf {\bibinfo {volume} {45}},\ \bibinfo
  {pages} {5} (\bibinfo {year} {2001})}\BibitemShut {NoStop}%
\bibitem [{\citenamefont {Fern{{\'a}}ndez-Delgado}\ \emph
  {et~al.}(2014)\citenamefont {Fern{{\'a}}ndez-Delgado}, \citenamefont
  {Cernadas}, \citenamefont {Barro},\ and\ \citenamefont
  {Amorim}}]{JMLR:v15:delgado14a}%
  \BibitemOpen
  \bibfield  {author} {\bibinfo {author} {\bibfnamefont {M.}~\bibnamefont
  {Fern{{\'a}}ndez-Delgado}}, \bibinfo {author} {\bibfnamefont
  {E.}~\bibnamefont {Cernadas}}, \bibinfo {author} {\bibfnamefont
  {S.}~\bibnamefont {Barro}},\ and\ \bibinfo {author} {\bibfnamefont
  {D.}~\bibnamefont {Amorim}},\ }\bibfield  {title} {\bibinfo {title} {Do we
  need hundreds of classifiers to solve real world classification problems?},\
  }\href {http://jmlr.org/papers/v15/delgado14a.html} {\bibfield  {journal}
  {\bibinfo  {journal} {Journal of Machine Learning Research}\ }\textbf
  {\bibinfo {volume} {15}},\ \bibinfo {pages} {3133} (\bibinfo {year}
  {2014})}\BibitemShut {NoStop}%
\bibitem [{\citenamefont {Hastie}\ \emph {et~al.}(2001)\citenamefont {Hastie},
  \citenamefont {Tibshirani},\ and\ \citenamefont
  {Friedman}}]{hastie01statisticallearning}%
  \BibitemOpen
  \bibfield  {author} {\bibinfo {author} {\bibfnamefont {T.}~\bibnamefont
  {Hastie}}, \bibinfo {author} {\bibfnamefont {R.}~\bibnamefont {Tibshirani}},\
  and\ \bibinfo {author} {\bibfnamefont {J.}~\bibnamefont {Friedman}},\
  }\href@noop {} {\emph {\bibinfo {title} {The Elements of Statistical
  Learning}}},\ Springer Series in Statistics\ (\bibinfo  {publisher} {Springer
  New York Inc.},\ \bibinfo {address} {New York, NY, USA},\ \bibinfo {year}
  {2001})\BibitemShut {NoStop}%
\bibitem [{\citenamefont {Breiman}(1996)}]{breiman96}%
  \BibitemOpen
  \bibfield  {author} {\bibinfo {author} {\bibfnamefont {L.}~\bibnamefont
  {Breiman}},\ }\bibfield  {title} {\bibinfo {title} {Bagging predictors},\
  }\href@noop {} {\bibfield  {journal} {\bibinfo  {journal} {Machine Learning}\
  }\textbf {\bibinfo {volume} {24}},\ \bibinfo {pages} {123} (\bibinfo {year}
  {1996})}\BibitemShut {NoStop}%
\end{thebibliography}
%

\appendix

\section{Recurrent Neural Networks} \label{sec:rnn}
In this section, we describe RNNs in additional detail. Our presentation is mostly standard and closely follows the treatment of \cite{jurafskyMartin24}.

RNNs are a class of models which employ a cycle within their network structure such that output of a unit depends in some manner on its own output at some previous step, so they have an underlying recursive structure \cite{rumelhart:errorpropnonote}. This structure allows RNNs to model sequences of data where information at the current index depends on information processed in the past. While the term RNNs refers to a general class of models which share this recursive pattern, in the ML literature, it is used to refer to a specific kind architecture with a simple cyclic pattern (sometimes also called Elman Networks) \cite{elman_rnn}.

\begin{figure}[th!]
    \centering
    \includegraphics[width=0.5\textwidth]{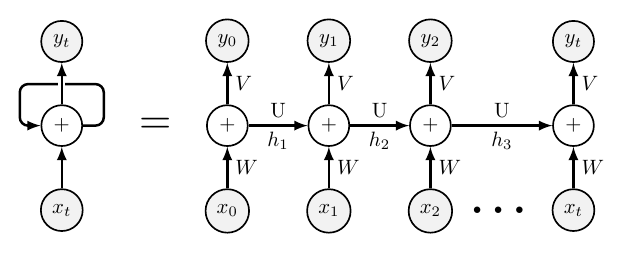}
    \caption{An illustration describing the basic architecture of a RNN. The model on the left can be thought of as \lq unrolling\rq\ itself in time.}
    \label{fig:elman}
\end{figure}

The model operates as follows (see Fig.~\ref{fig:elman}). Given an input vector $x_t$, where $t$ is an index over the sequence to be modelled, we first multiply $x_t$ with a weight matrix $W$ and pass the result through a non-linearity $g$ to produce values in a hidden layer $h_t$. However, unlike feed-forward networks we supplement this value with values from previous hidden layers $h_{t-1}$, multiplied by a weight matrix $U$ as well. This can be summarized as:
\begin{equation}
    h_t = g(Uh_{t-1} + Wx_t) . 
\end{equation}
Additionally, each unit of the RNN can produce an output $y_t$, which can be parameterized as follows:
\begin{equation}
    y_t = f(Vh_t), 
\end{equation}
where $f$ is usually taken to be a softmax function and $V$ is a weight matrix. This network is then trained through an analog of backpropagation which unrolls the sequence along its index referred to as Backpropagation through Time (BPTT) \cite{Robinson1987,Werbos1988,Mozer1995}.

In practice, it has been observed that such networks can be hard to train for longer sequence lengths. This is mainly due to the ``bottleneck" caused by the previous hidden state vector $h_{t-1}$ having to summarize \emph{all} of the relevant information at the current index inside a fixed size vector. Another difficulty encountered is due to repeated multiplications of small values in BPTT causing poor gradient flow to steps further back in time (referred to as the `vanishing gradient problem' \cite{Hochreiter:01book}.

LSTMs are a commonly used extension of RNNs to tackle the shortcomings pointed out above \cite{lstm}. The central idea is to gain more precise control over information being allowed to flow through the network. In addition to the hidden state, an additional so-called \lq memory cell\rq\ (also called the cell state) is computed, which, like the hidden state, allows the flow of information along the length of the sequence. The memory cell is designed such that information can be added to it or subtracted from it. 
We refer readers to \cite{goodfellow} for details of how this is done. 
\section{Baseline: Random Forests}
To justify our choice of employing LSTMs as the backbone of our proposed architecture, we provide a comparison  using Random Forests \cite{breiman2001random} for the encoder-only experiments described in Section \ref{sec_encoder} with. 
We choose to compare against Random Forests because they have been shown to have robust off-the-shelf performance across a wide variety of tasks \cite{JMLR:v15:delgado14a}. In this section we give a brief overview of Random Forests and a comparison with results in Table \ref{tbl:mse_encoder_only}, which justify the use of Neural Network based architectures in our subsequent experiments.

\subsection{Model Description}
Here we provide a brief overview of the Random Forest model, closely following the treatment in Ref.~\cite{hastie01statisticallearning}.
Decision trees are simple tree-based models that make decisions by recursively partitioning the input feature space into regions and then returning the label or output associated with that region. While decision trees are often robust to noisy features and exhibit a high degree of interpretability, they can grow to large depths, thereby overfitting the training set \cite{hastie01statisticallearning}. This effect is often addressed by learning a Random Forest, i.e., a collection (ensemble) of decision trees, each of which is constructed independently by training on a random subset of the data, in addition to using some random choices in the algorithm. This introduces diversity among the trees, preventing the model from becoming too dependent on any single feature. For regression tasks (i.e. the output is continuous-valued), as in this work, the final prediction is typically the average of the predictions made by each tree.

The combination of random sampling and majority voting (or averaging) is a technique known as Bagging (or \textbf{B}ootstrap \textbf{Agg}regat\textbf{ing})\cite{breiman96}. Random Forests use bagging to create an ensemble that is more robust (reducing uncertainty on the output) and less prone to overfitting compared to a single decision tree. Apart from their versatility and robustness, Random Forests provide insights into feature importance, helping to identify the most influential variables in the model.
\subsection{Training Setup}
Our aim is to compare the prediction accuracy for the parameter $\epsilon$ for a fixed $\Omega$ in the case of supervised learning between the Random Forest model and the model described in the main text. To do so, we use the same procedure as described in Sec.~\ref{sec_encoder} of the main text.
The training set contains averaged trajectories sampled randomly in groups of varying size $d$,
and the trajectories in the dataset correspond to $40$ values of $\epsilon$ evenly spaced on $[0,2)$ for a fixed $\Omega = 1.395$ radians/$\mu$s. The crucial hyperparameters for our method are the number of trees in the ensemble (set to 100) and the maximum depth of each tree (set to 25). These were found via 5-fold cross validation over the dataset.

A clear point of difference compared to the LSTM-based model is the form of the input given to the Random Forest model. Unlike the LSTM model, input to the Random Forest model \emph{must} be 1-dimensional. We use the concept of \lq dimensionality\rq\ to refer to the number of weak measurement values at each timestep. Our input is 2-dimensional since we have a weak measurement value for each qubit. This dichotomy requires us to \lq flatten\rq\ the input values into a single dimension. This can be done in two ways: either by keeping measurement values corresponding to a particular time step close together (column-major order) or by concatenating the values of the second qubit after \emph{all} the values of the first qubit (row-major order). We find that the choice of ordering makes little difference in our results as we demonstrate next. 
\subsection{Results}\label{sec:rf_results}
We present the impact of the choice of input order in our Random Forest model in Table~\ref{tbl:mse_rf_order_mse}. We find the difference in MSE values to be minimal. 
\begin{table}[!hbt]
\begin{tabular}{|c|c|c|c|}
\hline
$d$ & Best MSE & Median MSE & Mean MSE \\ \hline
2{,}000    & 2.35e-3, 2.63e-3 & 4.13e-3, 4.08e-3 & 4.41e-3, 4.52e-3 \\ \hline
4{,}000    & 1.23e-3, 1.31e-3 & 3.44e-3, 3.52e-3 & 3.69e-3, 3.70e-3 \\ \hline
8{,}000    & 7.49e-4, 7.34e-4 & 1.61e-3, 1.68e-3 & 1.86e-3, 1.86e-3 \\ \hline
16{,}000    & 3.36e-4, 3.26e-4 & 1.47e-3, 1.50e-3 & 1.65e-3, 1.71e-3 \\ \hline
$\infty$    & 2.34e-4, 3.48e-5 & 1.54e-3, 1.37e-3 & 1.75e-3, 1.79e-3 \\ \hline
\end{tabular}
\caption{Each cell shows MSEs for Random Forest models with row-major input flattening and column-major input flattening respectively.}
\label{tbl:mse_rf_order_mse}
\end{table}

\begin{figure}[th!]
    \centering
    \includegraphics[width=0.5\textwidth]{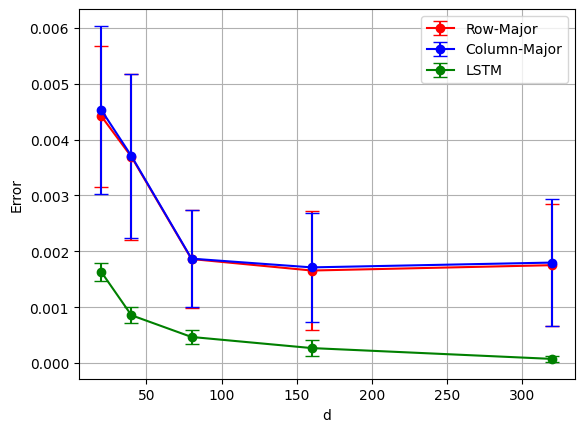}
    \caption{Mean MSEs and their 1 standard deviation error bars for the Random Forests model with different input orderings and the LSTM model.}
    \label{fig:error bars order}
\end{figure}

We now compare the results of the Random Forests model in column-major order to the LSTM model in Table~\ref{tbl:mse_baseline_encoder_only}.
We find about an order of magnitude difference in the MSEs achieved by the Random Forest model compared to the LSTM model. The trend of improved accuracy with increasing sizes of $d$, also observed in the LSTM results, continues to hold in this case. To get a better sense of the uncertainty, we also present the mean MSEs and their one standard deviation error bars for both our models in Fig.~\ref{fig:error bars order}. 
The fact that our choice of \lq flattening\rq\ does not affect overall performance of the Random Forest model suggests that the difference in performance between the Random Forest model and the LSTM model could be attributed to the LSTM-based model taking the sequential nature of the data into account versus the Random Forest model treating each trajectory as just a high-dimensional vector. We leave a thorough investigation of these issue to future work.
\begin{table}[!hbt]
\begin{tabular}{|c|c|c|c|}
\hline
$d$ & Best MSE & Median MSE & Mean MSE \\ \hline
2{,}000    & 2.63e-3, 1.47e-3 & 4.08e-3, 1.51e-3 & 4.52e-3, 1.63e-3 \\ \hline
4{,}000    & 1.31e-3, 7.37e-4 & 3.52e-3, 7.16e-4 & 3.70e-3, 8.56e-4 \\ \hline
8{,}000    & 7.34e-4, 4.39e-4 & 1.68e-3, 6.00e-4 & 1.86e-3, 4.67e-4 \\ \hline
16{,}000    & 3.26e-4, 1.83e-4 & 1.50e-3, 2.13e-4 & 1.71e-3, 2.68e-4 \\ \hline
$\infty$    & 3.48e-5, 8.11e-6 & 1.37e-3, 5.59e-5 & 1.79e-3, 7.41e-5 \\ \hline
\end{tabular}
\caption{Each cell shows the MSE pair for $\epsilon$ estimates on supervised training set using the encoder only. The first value is the MSE of the Random Forest model using column-major ordering, and second value corresponds to the MSE of the LSTM-based model.}
\label{tbl:mse_baseline_encoder_only}
\end{table}

We also provide a comparison of the number of parameters used in both our models in Table \ref{tbl:num_params}. Since the number of parameters in a Random Forest model depends on the depth of the trees, which in turn depend on the complexity of the dataset, we report the maximum parameters across all models obtained after training.

\begin{table}[!hbt]
\begin{tabular}{|c|c|c|c|}
\hline
Method & Number of parameters \\ \hline
LSTM   & 1,619,316 \\ \hline
Random Forest   & 129,380 \\ \hline
\end{tabular}
\caption{Number of parameters used in the two models studied.}
\label{tbl:num_params}
\end{table}

\end{document}